\documentclass[longauth]{aa}

\usepackage{graphicx}
\usepackage{color}
\usepackage[dvipsnames]{xcolor}
\usepackage{verbatim}
\usepackage[normalem]{ulem}
\usepackage{txfonts}
\usepackage[utf8]{inputenc}
\usepackage[pdftex,pagebackref,hyperindex=true,colorlinks=true,bookmarks=true,bookmarksopen=true,bookmarksnumbered=true,citecolor=OliveGreen]{hyperref}
\makeatletter
\renewcommand*\aa@pageof{, page \thepage{} of \pageref*{LastPage}}
\makeatother

\begin{document} 

   \title{MELCHIORS\thanks{The dataset is described in Appendix\,\ref{appendix}. It can be visualised and downloaded from \href{https://www.royer.se/melchiors.html}{www.royer.se/melchiors.html} and from the CDS, via anonymous ftp to \href{cdsarc.cds.unistra.fr}{cdsarc.cds.unistra.fr} (\href{130.79.128.5}{130.79.128.5}) or via \href{https://cdsarc.cds.unistra.fr/cgi-bin/qcat?J/A+A/}{https://cdsarc.cds.unistra.fr/cgi-bin/qcat?J/A+A/}.}}
   \subtitle{The Mercator Library of High Resolution Stellar Spectroscopy}



   \author{
           P. Royer\inst{\ref{kul}}\orcid{0000-0001-9341-2546}
          \and
          T. Merle\inst{\ref{rob},\ref{iaa}}\orcid{0000-0001-8253-1603}
          \and
          K. Dsilva\inst{\ref{kul},\ref{iaa}}\orcid{0000-0002-1476-9772} 
          \and
          S. Sekaran\inst{\ref{kul}}
          \and
          H. Van Winckel\inst{\ref{kul}}\orcid{0000-0001-5158-9327} 
          \and
          Y. Fr\'emat\inst{\ref{rob}}\orcid{0000-0002-4645-6017} 
          \and
          M. Van der Swaelmen\inst{\ref{oaa},\ref{iaa}}\orcid{0000-0002-1765-3521} 
          \and
          S. Gebruers\inst{\ref{kul}}\orcid{0000-0001-8426-4158}
          \and
          A. Tkachenko\inst{\ref{kul}}\orcid{0000-0003-0842-2374} 
          \and
          M. Laverick\inst{\ref{nz},\ref{kul}}\orcid{0000-0002-9220-2982} 
          \and
          M. Dirickx\inst{\ref{kul}} 
          \and
          G. Raskin\inst{\ref{kul}}\orcid{0000-0002-1323-9788} 
          \and
          H. Hensberge\inst{\ref{rob}}
          \and
          M. Abdul-Masih\inst{\ref{esochile},\ref{kul}}\orcid{0000-0001-6566-7568}
          \and
          B. Acke\inst{\ref{kul}} 
          \and
          M.L. Alonso\inst{\ref{kul}}
          \and
          S. Bandhu Mahato\inst{\ref{kul}}\orcid{0000-0002-1677-6474}
          \and
          P. G. Beck\inst{\ref{iac},\ref{ull},\ref{graz},\ref{kul}}\orcid{0000-0003-4745-2242}
          \and
          N. Behara\inst{\ref{iaa}}
          \and
          S. Bloemen\inst{\ref{rad},\ref{kul}}\orcid{0000-0002-6636-921X}
          \and
          B. Buysschaert\inst{\ref{kul}}
          \and
          N. Cox\inst{\ref{kul},\ref{nick}}\orcid{0000-0002-7926-4492} 
          \and
          J. Debosscher\inst{\ref{bira},\ref{kul}}
          \and
          P. De Cat\inst{\ref{rob}} 
          \and
          P. Degroote\inst{\ref{kul}}
          \and
          R. De Nutte\inst{\ref{rdw},\ref{kul}}\orcid{0009-0009-5807-5770} 
          \and
          K. De Smedt\inst{\ref{kul}}
          \and
          B. de Vries\inst{\ref{kul}}
          \and
          L. Dumortier\inst{\ref{rob}} 
          \and
          A. Escorza\inst{\ref{esochile},\ref{kul}}\orcid{0000-0003-3833-2513}
          \and
          K. Exter\inst{\ref{kul}} 
          \and
          S. Goriely\inst{\ref{iaa}} 
          \and
          N. Gorlova\inst{\ref{kul}}
          \and
          M. Hillen\inst{\ref{kul}}
          \and
          W. Homan\inst{\ref{kul}}
          \and
          A. Jorissen\inst{\ref{iaa}}
          \and
          D. Kamath\inst{\ref{kul}}
          \and
          M. Karjalainen\inst{\ref{cas}}\orcid{0000-0003-0751-3231}
          \and
          R. Karjalainen\inst{\ref{cas}}\orcid{0000-0002-2656-909X}
          \and
          P. Lampens\inst{\ref{rob}}\orcid{0000-0002-7034-4912}
          \and
          A. Lobel\inst{\ref{rob}}\orcid{0000-0001-5030-019X}
          \and
          R. Lombaert\inst{\ref{kul}}
          \and
          P. Marcos-Arenal\inst{\ref{kul}}\orcid{0000-0003-1549-9396}
          \and
          J. Menu\inst{\ref{kulitf},\ref{kul}}\orcid{0000-0001-6905-1840}
          \and
          F. Merges\inst{\ref{kul}}
          \and
          E. Moravveji\inst{\ref{kul}}
          \and
          P. Nemeth\inst{\ref{cas},\ref{hun},\ref{kul}}\orcid{0000-0003-0963-0239}
          \and
          P. Neyskens\inst{\ref{iaa}}
          \and
          R. Ostensen\inst{\ref{kul},\ref{rec}}
          \and
          P. I. P\'apics\inst{\ref{kul}}
          \and
          J. Perez\inst{\ref{kul}}
          \and
          S. Prins\inst{\ref{kul}}\orcid{0000-0002-6695-7759}
          \and
          S. Royer\inst{}
          \and
          A. Samadi-Ghadim\inst{\ref{gott},\ref{kul},\ref{rob}}
          \and
          H. Sana\inst{\ref{kul}}\orcid{0000-0001-6656-4130}
          \and
          A. Sans Fuentes\inst{\ref{kul}}
          \and
          S. Scaringi\inst{\ref{dur},\ref{kul}}\orcid{0000-0001-5387-7189}
          \and
          V. Schmid\inst{\ref{kul}}
          \and
          L. Siess\inst{\ref{iaa}} 
          \and
          C. Siopis\inst{\ref{iaa}} 
          \and
          K. Smolders\inst{\ref{kul}}
          \and
          S. Sodor\inst{\ref{rob}}
          \and
          A. Thoul\inst{\ref{ulg}}\orcid{0000-0002-8107-118X}
          \and
          S. Triana\inst{\ref{kul},\ref{iaa}}
          \and
          B. Vandenbussche\inst{\ref{kul}}\orcid{0000-0002-1368-3109}
          \and
          M. Van de Sande\inst{\ref{mvds},\ref{kul}}\orcid{0000-0001-9298-6265}
          \and
          G. Van De Steene\inst{\ref{rob}} 
          \and
          S. Van Eck\inst{\ref{iaa}}\orcid{0000-0003-0499-8608} 
          \and
          P.A.M. van Hoof\inst{\ref{rob}}\orcid{0000-0001-7490-0739} 
          \and
          A.J. Van Marle\inst{\ref{kul}}\orcid{0000-0002-2387-4515}
          \and
          T. Van Reeth\inst{\ref{kul}}\orcid{0000-0003-2771-1745}
          \and
          L. Vermeylen\inst{\ref{rob}}
          \and
          D. Volpi\inst{\ref{rob}}
          \and
          J. Vos\inst{\ref{cas},\ref{kul}}\orcid{0000-0001-6172-1272}
          \and
          C. Waelkens\inst{\ref{kul}}
          }

   \institute{Instituut voor Sterrenkunde, KU Leuven,
              Celestijnenlaan 200D bus 2401, 3001 Leuven, Belgium\label{kul}\\
              \email{pierre.royer@kuleuven.be}
        \and
        Royal Observatory of Belgium, Avenue Circulaire 3, 1180 Brussels, Belgium\label{rob}
        \and
        Institut d’Astronomie et d’Astrophysique, Universit\'e Libre de Bruxelles, CP 226 Boulevard du Triomphe, 1050 Bruxelles, Belgium\label{iaa}
        \and
        INAF - Osservatorio Astrofisico di Arcetri, Largo E. Fermi 5, 50125, Firenze, Italy\label{oaa}
        \and
        Centre for eResearch, The University of Auckland, Auckland 1010, New Zealand\label{nz}
        \and
        Royal Belgian Institute for Space Aeronomy (BIRA-IASB), 3 av. circulaire, B-1180 Bruxelles, Belgium\label{bira}
        \and
        Space sciences, Technologies and Astrophysics Research (STAR) Institute, Université de Liège, Allée du 6 Août 19C, Bat. B5C, B-4000 Liège, Belgium\label{ulg}
        \and
        ACRI-ST, Centre d’Etudes et de Recherche de Grasse (CERGA), 10 Av. Nicolas Copernic, 06130 Grasse, France\label{nick}
        \and
        Recogito AS, Storgaten 72, 8200 Fauske, Norway\label{rec}
        \and
        Department of Astrophysics/IMAPP, Radboud University, PO Box 9010, 6500 GL Nijmegen, The Netherlands\label{rad}
        \and
        Centre for Extragalactic Astronomy, Department of Physics, Durham University, South Road, Durham, DH1 3LE, UK\label{dur}
        \and
        Max-Planck-Institut für Sonnensystemforschung, 37077 Göttingen, Germany\label{gott}
        \and
        Astronomical Institute, Czech Academy of Sciences, 251 65 Ond\v{r}ejov, Czech Republic\label{cas}
        \and
        Astroserver.org, F\H{o} t\'{e}r 1, 8533 Malomsok, Hungary\label{hun}
        \and
        Redwire Space, Hogenakkerhoekstraat 9, 9150 Kruibeke, Belgium\label{rdw}
        \and
        Instituto de Astrof\'{\i}sica de Canarias, E-38200 La Laguna, Tenerife, Spain\label{iac}
        \and Departamento de Astrof\'{\i}sica, Universidad de La Laguna, E-38206 La Laguna, Tenerife, Spain \label{ull}
        \and Institut f\"ur Physik, Karl-Franzens Universit\"at Graz, Universit\"atsplatz 5/II, NAWI Graz, 8010 Graz, Austria \label{graz}
        \and
        Instituut voor Theoretische Fysica, KU Leuven, Celestijnenlaan 200D bus 2415, 3001 Leuven, Belgium\label{kulitf}
        \and
        European Southern Observatory, Alonso de C\'ordova 3107, Vitacura, Casilla 19001, Santiago de Chile, Chile\label{esochile}
        \and
        School of Physics and Astronomy, University of Leeds, Leeds LS2 9JT, UK\label{mvds}
        }

   \date{Received; accepted}

 
  \abstract
   {} 
   {Over the past decades, libraries of stellar spectra have been used in a large variety of science cases, including as sources of reference spectra for a given object or a given spectral type. Despite the existence of large libraries and the increasing number of projects of large-scale spectral surveys, there is to date only one very high-resolution spectral library offering spectra from a few hundred objects from the southern hemisphere (UVES-POP)
   . We aim to extend the sample, offering a finer coverage of effective temperatures and surface gravity with a uniform collection of spectra obtained in the northern hemisphere.  }
   {Between 2010 and 2020, we acquired several thousand echelle spectra of bright stars with the Mercator-HERMES spectrograph located in the Roque de Los Muchachos Observatory in La Palma, whose pipeline offers high-quality data reduction products. We have also developed methods to correct for the instrumental response in order to approach the true shape of the spectral continuum. Additionally, we have devised a normalisation process to provide a homogeneous normalisation of the full spectral range for most of the objects.}
   {We present a new spectral library consisting of 3256 spectra covering 2043 stars. It combines high signal-to-noise and high spectral resolution over the entire range of effective temperatures and luminosity classes. The spectra are presented in four versions: raw, corrected from the instrumental response, with and without correction from the atmospheric molecular absorption, and normalised (including the telluric correction).}
   
   {}


   \maketitle

\section{Introduction}
   
Theoretical libraries of stellar spectra offer noiseless data with a virtually infinite spectral resolution and an excellent coverage of the stellar parameter space. However, such libraries suffer from model-dependent systematic uncertainties, and they depend on the existence of reliable databases of atomic and molecular line lists and opacities~\citep{Jofre2019}.

Observational libraries, in contrast, do not rely on any assumption but often suffer from observational limitations in terms of spectral resolution, coverage of the parameter space, or signal-to-noise ratio.
In Table~\ref{tab_Libraries}, we list an overview of optical libraries that currently exist or are in the process of being acquired. (For an extended list covering older references, see~\cite{Sanchez06} and references therein. For a list of near-infrared libraries, see~\cite{Rayner09}.)



\begin{table*}[ht]
\caption{Optical spectral libraries. The term $N_*$ refers to the number of targets. The term $N_{sp}$ indicates the number of spectra when it is different from the number of targets. The wavelengths are expressed in nanometers.}
\label{tab_Libraries}
\centering
\begin{tabular}{l r r r r r l }     
\hline\hline       
Library  & $N_*$ & \hspace{-2mm}$N_{sp}$ & $\lambda_{min}$ & $\lambda_{max}$ & Resolution & Reference\\ 
\hline
    Gunn \& Strycker & 175 & & 313 & 1\,080 & 200 & \cite{GunnStryker1983}\\
    Silva \& Cornell & 72 & & 351 & 893 & 550 & \cite{SilvaCornell1992}\\
    James                       & 83 & & 350 & 750   & 800   & \cite{James13}\\ 
   Kitt-Peak                & 161 & & 351 & 742   & 1\,000  & \cite{Jacoby84} \\ 
   Pickles\tablefootmark{a} & 131 & & 115 & 2\,500  & 1\,000  & \cite{Pickles98} \\ 
    NGSL\tablefootmark{b}       & 378 & & 167 & 1\,025 & 1\,000 & \cite{Gregg06} \\ 
    SDSS-MaStar & 3\,321 & \hspace{-2mm}8\,646 & 362 & 1\,035 & 1\,800 & \cite{Yan19}\\ 
   STELIB                                   & 249 & & 320 & 930   & 2\,000  & \cite{LeBorgne2003} \\ 
        MILES\tablefootmark{c} & 985 & & 352 & 750 & 2\,000 & \cite{Falcon2011}\\ 
SDSS-BOSS\tablefootmark{d} & 324 & & 365 & 1\,020 & 2\,000 & \cite{Kesseli17} \\ 
Diaz et al. & 106 & & 790 & 910 & 2\,300 & \cite{Diaz1989}\\ 
    MUSE library & 35 & & 480 & 930 & 3\,000 & \cite{Ivanov19}\\ 
        Indo-US & 1\,273 & & 346 & 946 & 5\,000 & \cite{Valdes04}\\ 
   XSL\tablefootmark{e}     & 683 & \hspace{-2mm}830 & 350 & 2\,480  & 10\,000 & \cite{Gonneau20,Verro22}\\ 
   ELODIE                       & 1\,388 & \hspace{-2mm}1\,962 & 390 & 680   & 42\,000 & \cite{Prugniel01,Prugniel04,Prugniel07}\\ 
    FGKM library & 404 & & 499 & 641 & 60\,000&\cite{Yee2017}\\ 
 Gaia-FGK\tablefootmark{b,f} & 34 & \hspace{-2mm}71 & 300 & 1\,020 & 80\,000 & \cite{Blanco14}\\
   UVES-POP\tablefootmark{g}                 & 394 & & 304 & 1\,040  & 80\,000 & \cite{Bagnulo03} \\  
MELCHIORS & 2\,043 &  \hspace{-2mm}3\,256 & 380 & 900 & 85\,000 & This work\\
\hline
\end{tabular}
\tablefoot{
\tablefoottext{a}{Pickles: Spectra assembled from various sources.}
\tablefoottext{b}{NGSL, Gaia-FGK: High signal-to-noise.}
\tablefoottext{c}{MILES: Spectrophotometry.}
\tablefoottext{d}{SDSS-BOSS: Spectral templates assembled from low-resolution spectra extending down to spectral type L.}
\tablefoottext{e}{X-shooter Spectral Library: Data release 3.}
\tablefoottext{f}{Gaia-FGK: Limited to spectral types F, G, K.}
\tablefoottext{g}{UVES-POP's spectral resolution reaches 120\,000 over part of the range.}
}
\end{table*}

Different spectral surveys target different goals, and this is reflected in the properties of the final libraries. Despite its limited size (161 spectra), resolution ($\sim\!\!1000$), S/N ($\sim\!\!100$), and wavelength coverage (351--742\,nm), the library of \cite{Jacoby84}, for instance, is still highly cited nowadays, especially thanks to its large coverage of the Hertzsprung-Russel Diagram (HRD). 

Over the past decades, much larger and systematic efforts have been developed on the basis of much more extensive spectral libraries. As an example, SDSS-APOGEE aims at studying the galactic structure and evolution via the chemical peculiarities of various populations of stars. The survey therefore covered 100\,000 high S/N spectra in the H-Band at a resolution of 20\,000 \citep{Allende08}. In the visible, another example is the Gaia-ESO Survey \citep{Gilmore12,Gilmore22,Randich13,Randich2022}, which  targeted 100\,000 stars in all the components of the Milky Way at medium and high resolution with GIRAFFE and UVES spectrographs. 
Other surveys, such as the Gaia FGK Benchmark Stars project \citep{Blanco14}, use the high-quality spectra to provide the community with benchmarks for parameter determinations and abundance derivations \citep{Jofre17}. Projects like BRASS \citep{Laverick18a,Laverick18b} and \cite{Ruffoni14} exploit high-quality data to attack the uncertainties in atomic data parameters (i.e. part of the systematic uncertainties underlying the production of theoretical spectra). 

While huge spectral libraries are becoming available and offer invaluable input for statistical analyses, they come at the price of mandatory trade-offs in the observable parameter space. Huge numbers of targets can only be observed with reduced spectral resolution or (but often and) reduced S/N. Moreover, the individual targets in these types of observations cannot benefit from the same care in data reduction as is possible for smaller-scale projects.

While this illustrates the need for such spectral libraries, a quick look at Table~\ref{tab_Libraries} also shows that a large collection of spectra that has high-resolution (significantly larger than 30\,000), has a high S/N, widely covers the HRD and offers high-quality data products is still missing. A noticeable exception is UVES-POP \citep{Bagnulo03}, a filler programme on the ESO-VLT instrument UVES, which delivered a database consisting of 394 spectra of high-spectral resolution and high S/N. The data reduction pipeline applied to these echelle spectra was unfortunately sub-optimal in some aspects (e.g. order matching), resulting in imperfect representations of the spectral continua.
These spectra were nevertheless instrumental in studies covering a wide range of applications (see the 350 citations to the UVES-POP Messenger article), re-enforcing the need for such high-quality, high-resolution, and high S/N spectral libraries.

From 2010 to 2020, we conducted a spectral survey of bright stars from the northern hemisphere as a filler programme with the Mercator-HERMES spectrograph \citep{Raskin2011} attached to the Mercator telescope at the Roque de Los Muchachos Observatory on the island of La Palma, Spain. The resulting library combines high-spectral resolution (85\,000) and high S/Ns. It was also designed for densely sampling the HRD using optically bright objects. 

As our aim is to also offer high-quality data products, including automatically normalised spectra for quick-look analyses (while being science ready in most circumstances), the present database gathers some unique properties when compared to the other libraries reported in Table~\ref{tab_Libraries} in terms of coverage of the HRD, spectral resolution, S/N, high-quality data reduction, and sample size. The potential uses of such an empirical library include population synthesis for galactic and extra-galactic studies, references for stellar atmosphere models, spectral classification, and studies of binaries and outreach. It may also constitute a nice reference sample to train neural networks for data-driven machine learning applications.

We present the target selection and observations in Section~\ref{sec:selec} and the data processing steps in Section~\ref{sec:data_proc}, including the principle of the instrumental response correction and of the automatic spectral normalisation. The resulting database is presented in Section~\ref{sec:results}, along with the distribution on the sky, in the HRD or in S/N. We present some brief conclusions in Section~\ref{sec:conclusions}.

\section{Target selection and observations}
\label{sec:selec}

Our prime goal was to sample the HRD as densely as possible while minimising the required exposure time. It was also clear from the beginning that assembling a large spectral database could only be possible in the context of a filler programme. Therefore, our targets would have to be bright and observable even in poor seeing conditions as well as through thin clouds. Finally, we had to ascertain a good distribution in right ascension in order to offer possibilities for short but useful observations whenever gaps in the regular observing programmes would appear. Consequently, the original target selection was based on simple criteria.

We started compiling our sample by collecting all objects at a declination higher than $-30$ degrees, due to the latitude of the observatory, and with an apparent magnitude of $V < 8$, given the filler nature of the programme. We retrieved the corresponding 34263 objects from Simbad and rejected the non-stellar objects (360) as well as those bearing a bad V-band magnitude quality flag (flag `E'; 402 objects), leaving 33511 objects (10 entries were rejected for both criteria).

We then sorted the objects by spectral type, and out of the 2006 unique spectral type strings, we selected the well-behaved strings (i.e. rejecting classifications such as `K2/K3IIICN...' or `$\sim$'). Our goal was to build a library of reference spectra from which spectral templates could potentially be derived, but we did not aim at maximising diversity. Hence, this selection automatically removed the stars known as emission line stars as well as those known to be magnetic and/or chemically peculiar. This first selection left us with 28247 objects. In order to bring this number down to a reasonable requirement in terms of telescope time, we then limited the input catalogue to the ten brightest objects for each type, sub-type, and luminosity class (e.g. F8V or A0III), providing a selection of 2178 targets. 

The exposure time calculator of the HERMES spectrograph was used in combination with Simbad's B and V magnitudes in order to estimate the required exposure times, targeting a S/N of 200. The filler nature of the programme made it so that the final S/N was often slightly inferior when the observer was not tracking the S/N in real time or when the exposure times were so short that this was not possible in practice.

The observations were acquired over 513 different nights between April 7, 2010, and February 12, 2020. The observations acquired prior to September 28, 2010, were later excluded, as no reliable meteorological data exist for that period, which hampered the correction for the molecular absorption in the corresponding 125 spectra (section~\ref{sec:atm_corr}).

In April 2013, at the request of the BRITE consortium (K. Zwintz, priv. comm.), 95 objects brighter than $m_v=4$ and missing from the original list were added to the sample. Seventy four O-type stars were also re-introduced at the same time to compensate for their scarcity in the original catalogue as much as possible.

In May 2014, a majority of the targets had been observed already, and the complete catalogue was re-submitted to the scheduler of Mercator-HERMES. This would restore the availability of filler targets all year long while providing interesting information on the variability or binarity for our targets. For this resubmission, a few entries had to be manually edited, to correct for inaccurate information from Simbad. In particular, three duplicate entries were removed, leaving a total of 2360 objects. 

Not all objects in the catalogue were actually observed. A bias exists against the faintest targets. These are indeed more difficult to schedule, especially in the context of a filler programme. We find it is worth noting, however, that a handful of objects that were not in the original catalogue were actually observed as part of this programme at the discretion of the observers on duty during the observation shift. 

\begin{table}[t]
\caption{Wavelength ranges and the respective molecules included in the fitting process with {\tt molecfit.}}
\begin{tabular}{ccl}
\hline \hline
Wavelength start (\AA) & Wavelength end (\AA) & Molecule \\
\hline
6274 & 6324 & O$_2$     \\
6850 & 7074 & O$_2$ (B band)     \\
7227 & 7269 & H$_2$O    \\
7589 & 7714 & O$_2$ (A band)    \\
8101 & 8344 & H$_2$O    \\
8940 & 8971 & H$_2$O    \\
\hline
\end{tabular}
\label{tab:molecfitranges}
\end{table}

\begin{figure}
\centering
\includegraphics[width=\hsize]{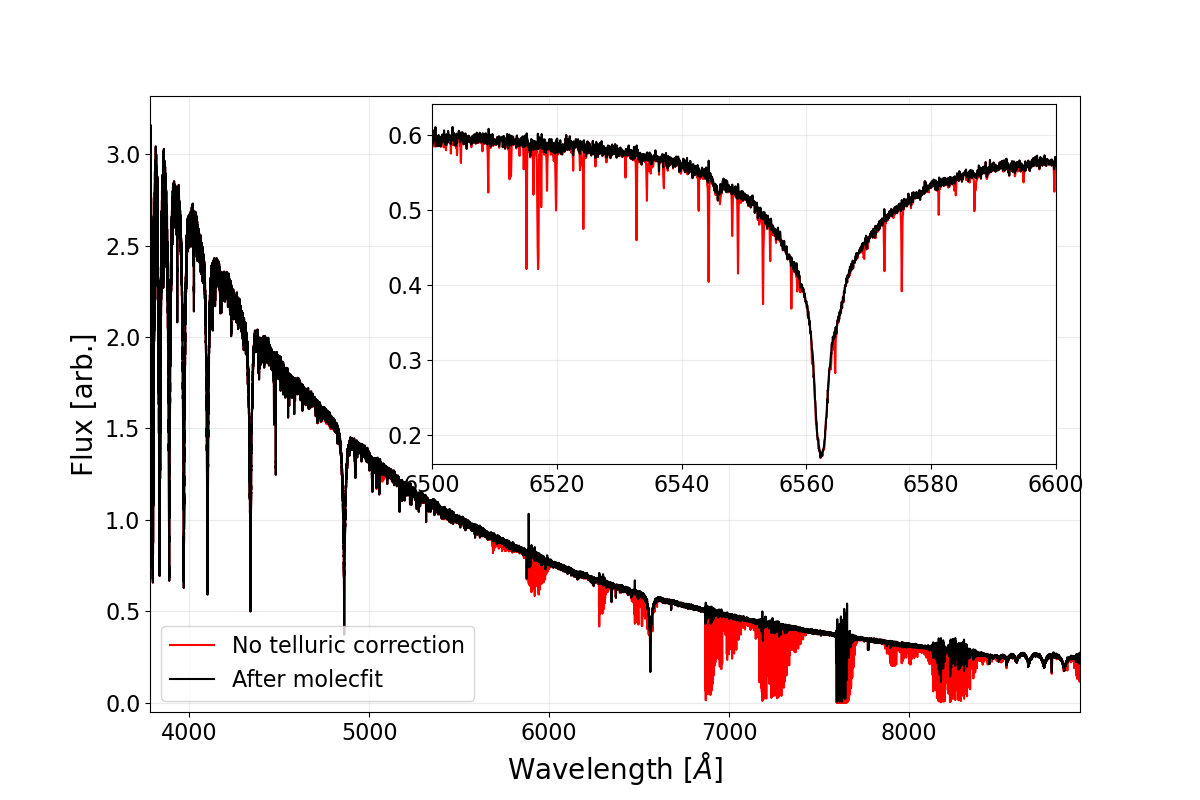}
     \caption{Effect of the telluric correction on a spectrum from HD\,19356 (B8V). In red: the spectrum before the correction. In black: the spectrum after  correction with {\tt molecfit}. The insert shows the region around $H\alpha$. Whereas weak lines are usually very well corrected, large correction residuals do appear at the level of saturated telluric lines.}
     \label{fig_mfitcorr}
\end{figure}

\section{Data processing}
\label{sec:data_proc}

At the entrance of the telescope, any stellar spectrum S can be decomposed in a number of factors: 
\begin{equation}
    S = C \cdot\ N \cdot\ A^e \cdot\ A^m,
\end{equation}
where $A^e$ is the atmospheric extinction, $A^m$ is the molecular absorption by the Earth's atmosphere (telluric lines), and $S^0 = C\,\cdot\,N$ is the stellar spectrum (including the interstellar reddening). Finally, $C$ represents the stellar continuum, that is, the ratio between $S^0$ and the normalised spectrum $N$. The spectrum $S$ contains the effect of the interstellar absorption, which we do not correct and is consequently included in $C$. In the present work, we aim at providing three types of data products: the response-corrected spectra before and after correction for the telluric lines (i.e. $S^{atm}_{OBJ} \triangleq C\,\cdot\,N\,\cdot\,A^m$ and $S^0_{OBJ} \triangleq C\,\cdot\,N$, respectively) and the normalised spectra $N$ of the telluric-corrected spectrum.

To this aim, we considered three types of measurements: observations of scientific targets, observations of spectrophotometric calibrators, and flat-field exposures. After subtraction of the electronic offset, these measurements can be expressed as:
    \begin{align}
    \label{eq:obj}
    f_{obj} &= S_{OBJ}\,\cdot\,\Re',\\
    \label{eq:cal}
    f_{cal} &= S_{CAL} \cdot\, \Re', \\
    \label{eq:ff}
    ff_{raw} &= FF_{SED} \cdot\, \Re'.
    \end{align}
The left-hand sides of these equations represent the three types of measurements, and $FF_{SED}$ expresses the global shape of the flat-field signal imposed by the spectrum of the lamps and the filters used for these measurements. $\Re' = \Re\,\cdot\,B\,\cdot\,P$ describes the instrumental response in which $B$ represents the blaze function driving the response inside every order of the echelle spectrometer, $P$ is the pixel response non-uniformity (PRNU), and $\Re$ (hereafter called `the response') captures the remaining instrumental effects. The complex shape of the response $\Re$ is determined by the flat-field procedure in the HERMES instrument, which relies on the combination of a high-powered tungsten lamp with red blocking filters to obtain enough blue photons and a low-powered tungsten lamp for the red spectral orders \citep{Raskin2011}.

\begin{table*}
\setlength{\tabcolsep}{8pt}
\caption{Calibration stars used for the instrumental response correction, including their observation frequency and the astrophysical parameters used as input to generate our SED models. Errors on the astrophysical parameters derived from atmospheric modelling in \textsc{gssp} are on the order of 100 K in $T_{\text{eff}}$, 0.1 dex in $\log{g}$, 1 km/s in $v\sin{i}$, 0.1 km/s in $v_{\text{mic}}$, and 0.1 dex in $\mathrm{[M/H]}$.
}
\label{tab:P2}      
\centering                          
\begin{tabular}{l r c r c r r}        
\hline\hline                 
Target & $T_{\text{eff}}$ [K] & $\log{g}\,[\text{dex}]$ & $v\sin{i}\,[\text{km/s}]$ & $v_{\text{mic}}$ [km/s] & $\mathrm{[M/H]}$ & \#nights \\    
\hline                        
HD 36267 & 15250 & 4.2 & 127 & 0.6 & 0.0 & 29 \\ 
HD 152614 & 11750 & 3.9 & 109 & 1.5 & 0.0 & 40 \\
HD 149212 & 10500 & 3.8 & 157 & 0.0 & 0.4 & 10 \\
HD 14055 & 10250 & 4.3 & 235 & 2.6 & $-0.1$ & 58 \\
HD 87887 & 10250 & 3.7 & 22 & 1.2 & 0.1 & 26\\
HD 42818 & 10000 & 4.2 & 275 & 2.5 & 0.1 & 18\\
HD 214994 & 9500 & 3.6 & 15 & 2.1 & 0.2 & 6\\
HD 46300\tablefootmark{a} & 9500 & 2.0 & 100 & 2.0 & 0.0 & 36 \\   
HD 118098 & 8300 & 3.7 & 213 & 2.3 & $-0.2$ & 66 \\
GSC4293-0432 & 8000 & 4.0 & 40 & 3.1 & 0.3 & 2 \\
HD 56169 & 7900 & 2.7 & 216 & 3.3 & $-0.6$ & 27\\
HD 184006 & 7800 & 3.0 & 245 & 3.6 & $-0.4$ & 46 \\
HD 147449 & 7100 & 4.0 & 79 & 2.9 & 0.0 & 17\\
HD 84937 & 6400 & 3.8 & 5 & 1.7 & $-2.1$ & 12\\
\hline                                   
\end{tabular}
\tablefoot{
\tablefoottext{a}{The model for this object was taken from the synthetic library of \citet{Munari05}.}
}
\end{table*}

The response $\Re$ is both wavelength and time dependent. The wavelength dependence describes the relative spectral response function (RSRF), and the time dependence takes into account possible variations of the environmental conditions (temperature, atmospheric pressure, and humidity), instrument ageing, and (in our case) residuals with respect to the average atmospheric extinction (see~\ref{sec:atm_corr} below). Formally, the response encapsulates the flux calibration coefficient, which we disregard in this work, as our spectra are not calibrated in flux. 

The flat-field exposures are usually acquired at the beginning and the end of the night. The celestial calibrators are objects whose spectra are known from accurate modelling. One such object was observed every night whenever possible. We used the celestial calibrators to correct for the instrumental response (Section~\ref{sec:responsecorr}).

\subsection{Pipeline}
The raw 2D echelle spectra were first processed with the standard HERMES pipeline, which includes bias correction, background subtraction, flat-field correction, cosmic-ray clipping, order merging, and wavelength calibration, including barycentric correction \citep{Raskin2011}. 
The wavelength calibration is based on a spectral mask containing 1600 carefully selected isolated Th, Ar, and Ne lines. This mask contains the 2D pixel coordinates and identification of all lines. During the wavelength calibration, the positions of these lines were carefully measured in the 2D extracted pixel-order space.  Except for the bluest and the two reddest orders, the wavelength calibration per order is based on 20 to 50 spectral lines. A 2D polynomial fit of the order (8,\,8) in pixel-order space determines the wavelength calibration. The inter-order calibration is typically better than 50~m~s$^{-1}$, as evaluated from a cross-correlation measurement using a ThArNe spectral mask on a ThArNe extracted file.  For order 40, which is the reddest spectral order, we had to derive the calibration from only one Ne line, and the wavelength calibration is significantly poorer.

We measured wavelength calibration frames typically just before and after the astronomical night, with one additional calibration in the middle of the night. Since the spectrograph is not mounted in a vacuum tank, it is subject to atmospheric pressure variations. Nevertheless, thanks to the control and stability of the temperature within the spectrograph room, the velocity stability is good, with a 1 $\sigma$ level of 60~m~s$^{-1}$ on 3432 measured radial velocity standards over a period of 10 years. For the purpose of line identification and the spectral characterisation of stellar sources, this wavelength stability and calibration homogeneity is excellent.

The flat-field correction is a mandatory step to remove the PRNU and the effect of the blaze function in every order prior to order merging. Dividing by the flat-field nevertheless artificially imposes the shape of $1 / FF_{SED}$ onto every spectrum thus corrected. The output of the pipeline hence consists of 1D spectra still influenced by the effects of the Earth's atmosphere and the newly introduced shape of the flat-field. 

\subsection{Atmospheric extinction and telluric correction}
\label{sec:atm_corr}
To correct for the atmospheric extinction $A^e$, a wavelength and airmass dependent atmospheric correction was applied following the tabulated values for the Roque de Los Muchachos Observatory on the island of La Palma.\footnote{\href{https://www.ing.iac.es/Astronomy/observing/manuals/html\_manuals\\/general/obs\_guide/node293.html}{https://www.ing.iac.es/Astronomy/observing/manuals/html\_manuals\\/general/obs\_guide/node293.html}} This resulted in an underestimation of the extinction for part of the observations (e.g. those performed under thin clouds). We note however that despite the filler nature of our programme, a fair fraction of the observations were executed under average to good weather conditions, as the short exposure times and extensive coverage of the sky offered easy-to-schedule solutions to fill short gaps in the regular observing programmes.

To correct for the telluric contamination $A^m$, we used the publicly available tool {\tt molecfit} \citep{2015SmetteMolecfit,2015KauschMolecfit}. By combining essential input from the Mercator weather station with a molecular line database and a radiative transfer code, {\tt molecfit} models the instrument line spread function and atmospheric content at the time of the observation.\footnote{Due to the absence of meteorological data records prior to September 28, 2010, the spectra obtained before that date were excluded to ensure the homogeneity of the dataset.} The wavelength ranges and molecules included in the fitting procedure are listed in Table\,\ref{tab:molecfitranges}. For every region included in the fit, the continuum is estimated locally using a first-order polynomial. The {\tt molecfit} tool also corrects for any minor wavelength shifts by fitting the wavelength with a second-order Chebyshev polynomial. The differences with the original wavelength calibration were found to be negligible ($\leq\!100$~m~s$^{-1}$). Hence, we kept the original wavelength calibration in the final products.

After correction, the spectra showed residuals of $\sim\!\!5\%$ of the continuum in the telluric bands, with the exception of the saturated lines (Fig.\,\ref{fig_mfitcorr}). While fitting saturated lines, the division of two small numbers often results in large residuals (the oxygen band around $7600\,$\AA\, is often impacted this way). In the region above $8950\,$\AA\,, we observed `P\,Cygni'-shaped residuals due to a small imperfection of the wavelength calibration solution at the very edge of the spectra. In order for the user to estimate the impact of {\tt molecfit} on the data, we provide both versions of the spectra, that is,  before and after correction.

\begin{figure}
\centering
\includegraphics[width=\hsize]{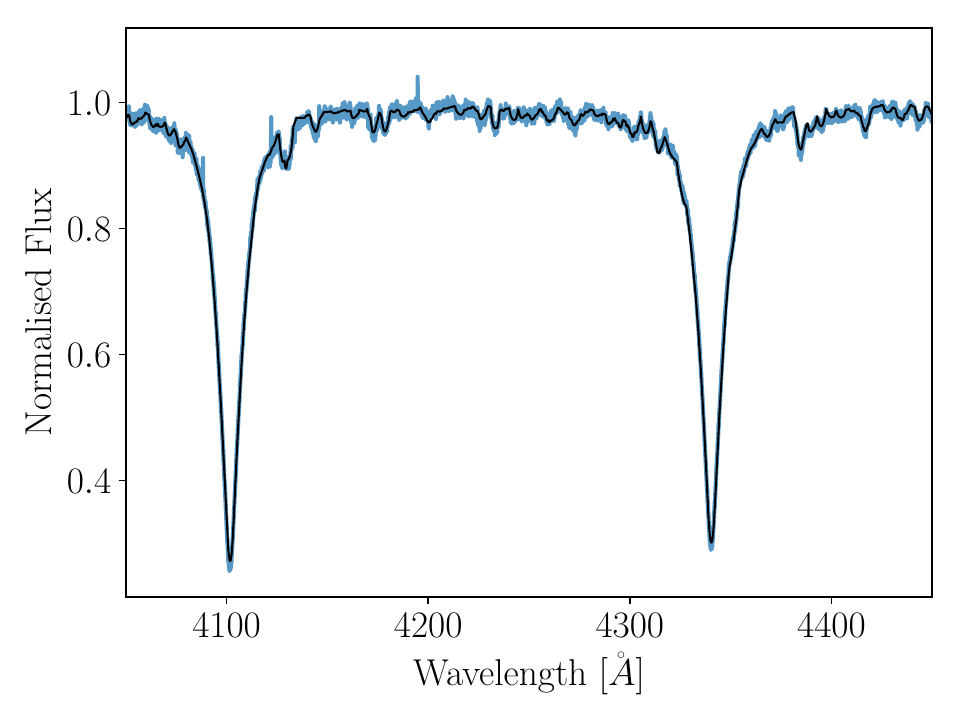}
     \caption{Best-fitting normalised spectroscopic model (in black) of the highest-S/N observed spectrum of HD 149212 (in blue), focussing on the wavelength region between the H$_{\gamma}$ and H$_{\beta}$.}
     \label{fig_HD149212_bestfit}
\end{figure}

\subsection{Response correction}
\label{sec:responsecorr}

At this stage, the spectra could be expressed as:
    \begin{align}
    \label{eq:divff0obj}
    f_{obj}^{0,FF} &= \frac{S^0_{OBJ}\,\cdot\,\Re}{ff},\\
    \label{eq:divff0cal}
    f_{cal}^{0,FF} &= \frac{S^0_{CAL}\,\cdot\,\Re}{ff},
    \end{align}
where `$FF$' and `$0$' indicate that the flat-field and atmospheric corrections have been applied and $ff \triangleq FF_{SED}\,\cdot\,\Re$. If we define an effective instrumental response $\hat{\Re}$ combining the system response with the effect of the flat-field process, that is, 
\begin{equation}
    \label{eqn:hatre}
    \hat{\Re} \triangleq \frac{\Re}{ff} = \frac{f^{0,FF}_{cal}}{S^0_{CAL}},
\end{equation}
we can express the final spectrum corrected from the instrumental response, the effects induced by the flat-field correction, and the effects from the atmosphere as:
\begin{equation}
\label{eqn:Sfinal}
S^0_{OBJ} = \frac{f^{0,FF}_{obj}}{\hat{\Re}}.
\end{equation}

\begin{figure}[t]
\centering
\includegraphics[width=\hsize]{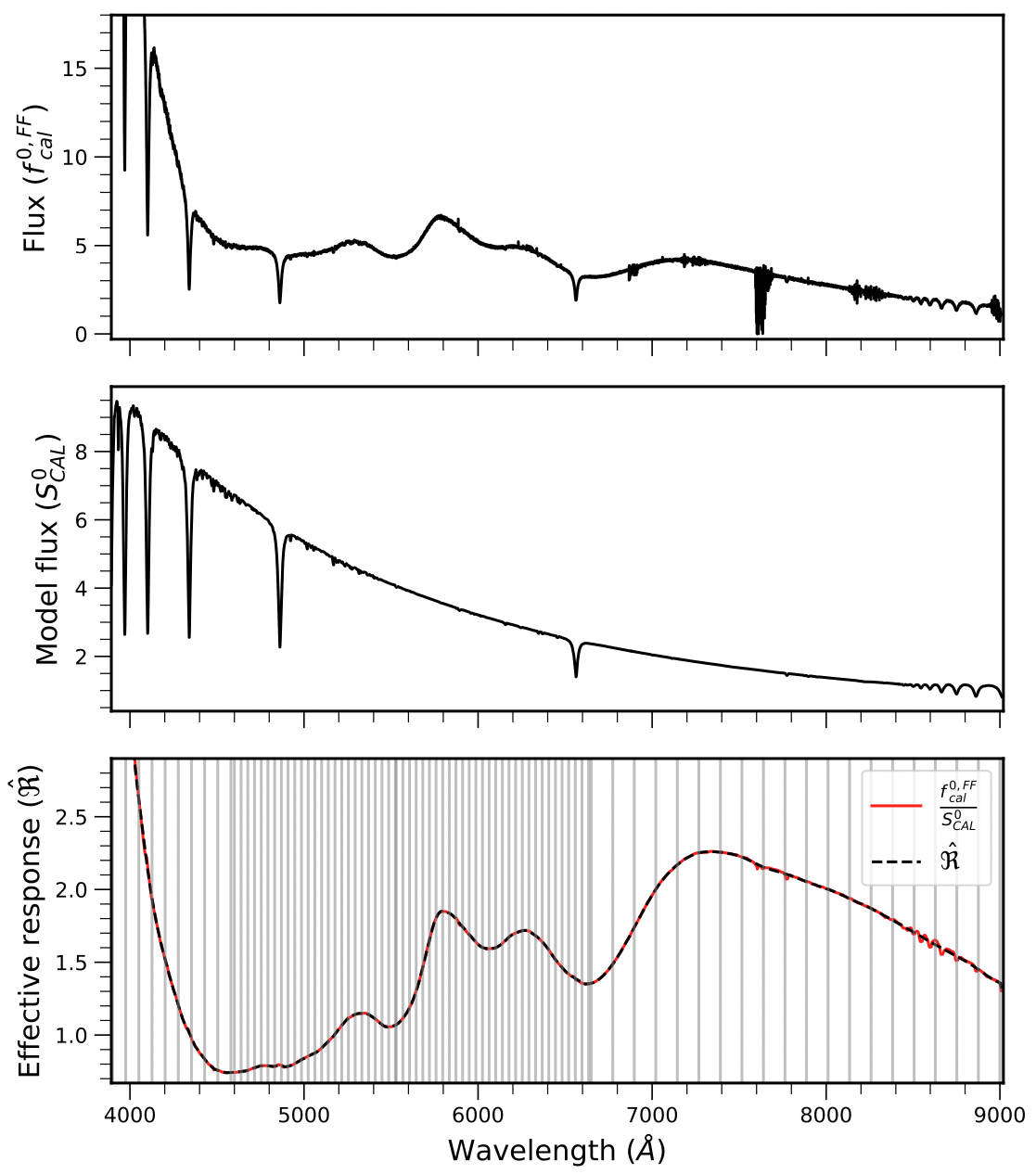}
     \caption{The derivation of the effective response for a night (eqn. \ref{eqn:hatre}). The spectrophotometric calibrator star used in this example is HD 42818. Top: the spectrum corrected for flat-field and atmospheric effects ($ f_{cal}^{0,FF}$). Middle: Model SED used to derive the response ($S^0_{CAL}$). Bottom: The derived effective response in red which is fit with a spline in black ($\hat{\Re}$). The knot points used to fit the spline are shown with vertical lines. }
     \label{fig_respDerivation}
\end{figure}

\begin{figure}[t]
\centering
\includegraphics[width=\hsize]{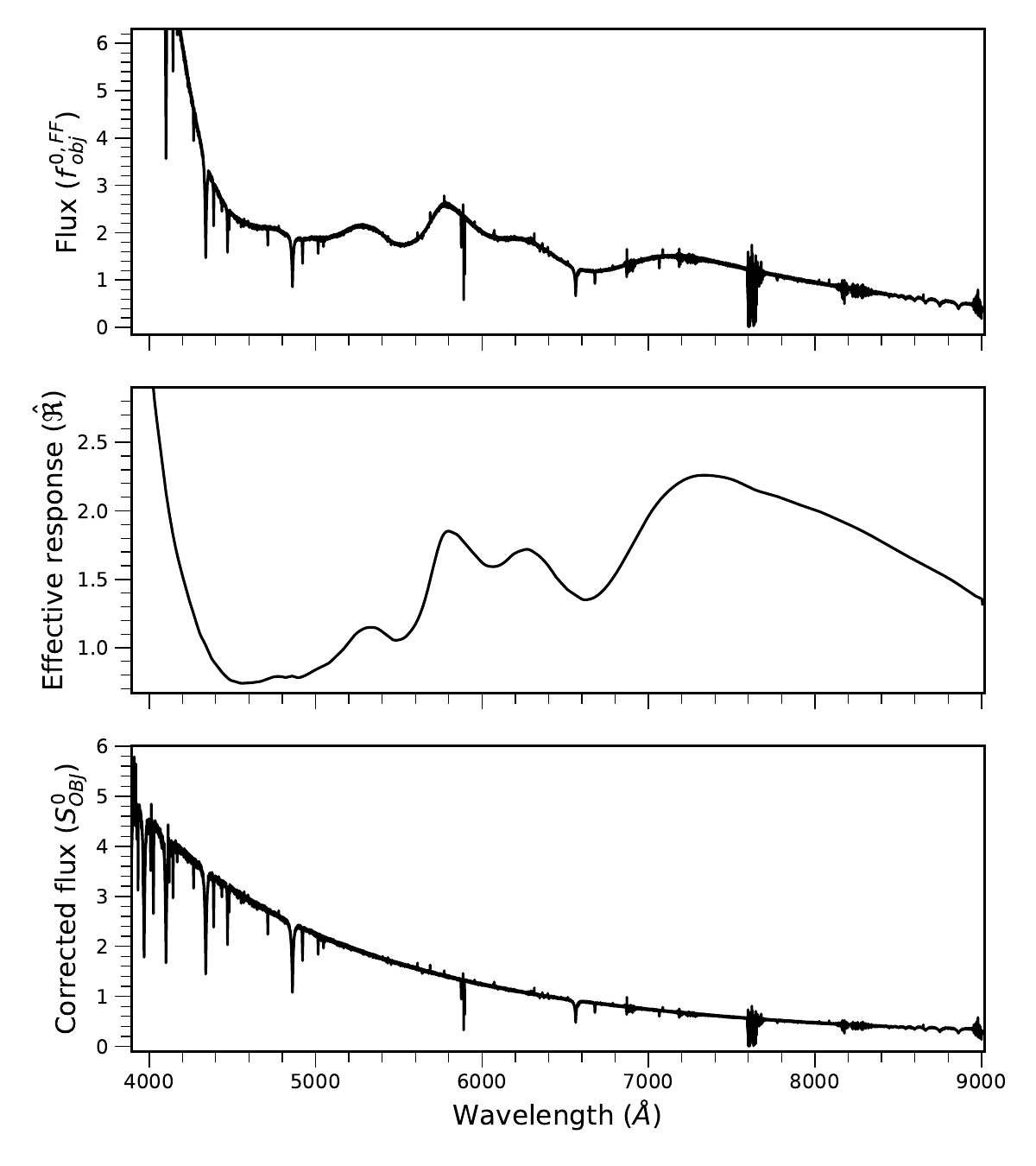}
     \caption{Instrumental response correction for a spectrum of the scientific target HD\,3379 (B2.5IV). Top: Spectrum corrected for flat-field and atmospheric effects ($ f_{obj}^{0,FF}$). Middle: Effective response ($\hat{\Re}$) for the same night. Bottom: Response-corrected spectrum ($S^0_{OBJ}$).}
     \label{fig_FullCorrHD3379}
\end{figure}

One calibration star was observed every night when possible as part of the normal calibration set of HERMES, aiming at a daily determination of $\hat{\Re}$. These targets were carefully selected for this purpose, as they are well documented in the literature and have reliable models established by a series of independent publications.  All of these targets were analysed with the LTE spectral-fitting code \textsc{gssp} \citep{Tkachenko2015}. The \textsc{gssp} code generates synthetic spectra using the \textsc{SynthV} radiative transfer code \citep{tsymbal1996} combined with a grid of atmospheric models from the \textsc{LLmodels} code \citep{Shulyak2004}. These synthetic spectra were then fitted to the highest-S/N observed spectrum for each star in the HERMES archive. The atmospheric parameters $T_{\text{eff}}$, log $g$, projected rotational velocity ($v$~sin~$i$), microturbulent velocity ($v_{\text{mic}}$), and the metallicity ($\mathrm{[M/H]}$) as well as their corresponding uncertainties were then determined from the distribution of $\chi^{2}$ values of the fit of each synthetic spectrum to the observed spectrum. We then used the parameters in the \textsc{LLmodels} grid closest to our best-fitting values as the input to generate the spectral energy distribution models of our calibrators. An example of the fit is shown in Fig.~\ref{fig_HD149212_bestfit}. The list of calibrators, the astrophysical parameters of the selected models, and the number of nights they were observed are presented in Table~\ref{tab:P2}.

\begin{figure}[ht]
\centering
\includegraphics[width=\hsize]{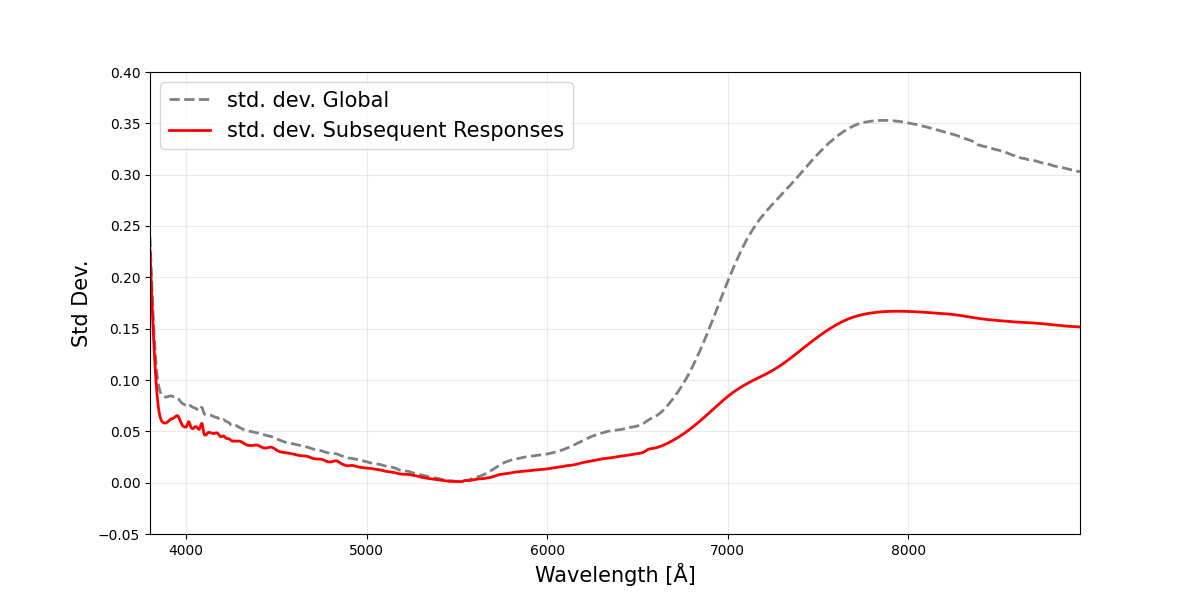}
     \caption{Variability of the instrumental response. Red: Standard deviation of the ratio between subsequent responses derived for our programme indicative of the additional uncertainty on the low-frequency components of the continuum for the stars with {\sc stdnight}=0. Grey: Standard deviation computed over all responses, regardless of their acquisition time. (See text for details.)}
     \label{fig_stddev_resp}
\end{figure}

\begin{figure*}[bht]
    \centering
    \includegraphics[width=\linewidth]{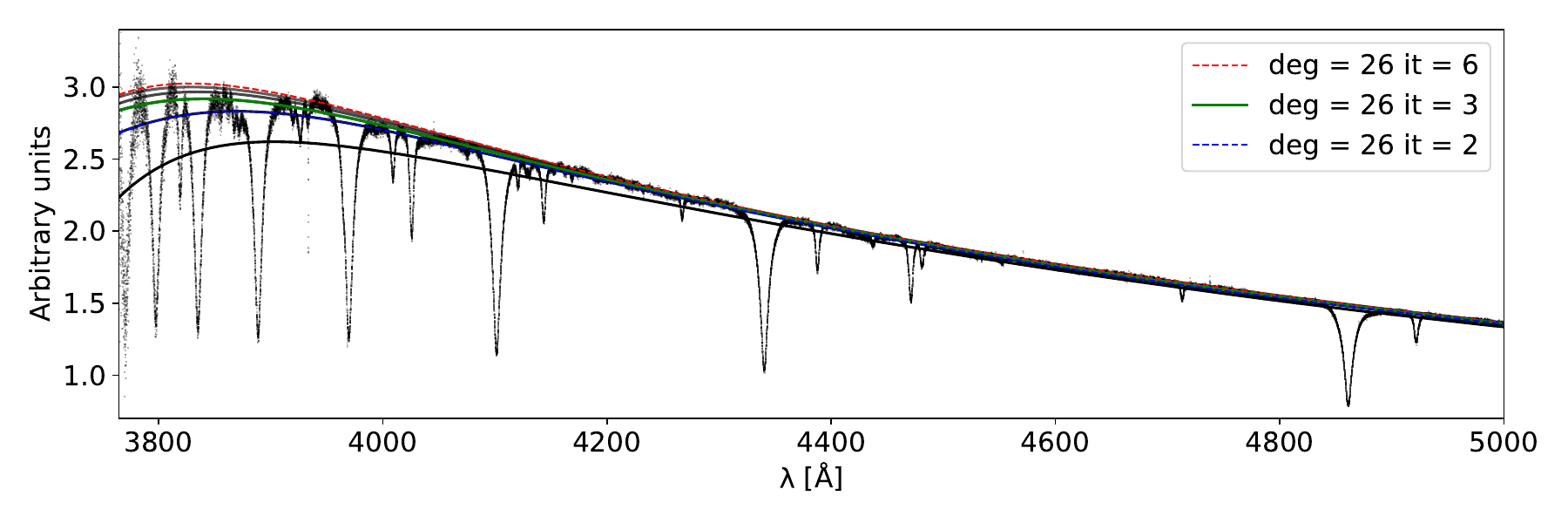}
    \includegraphics[width=\linewidth]{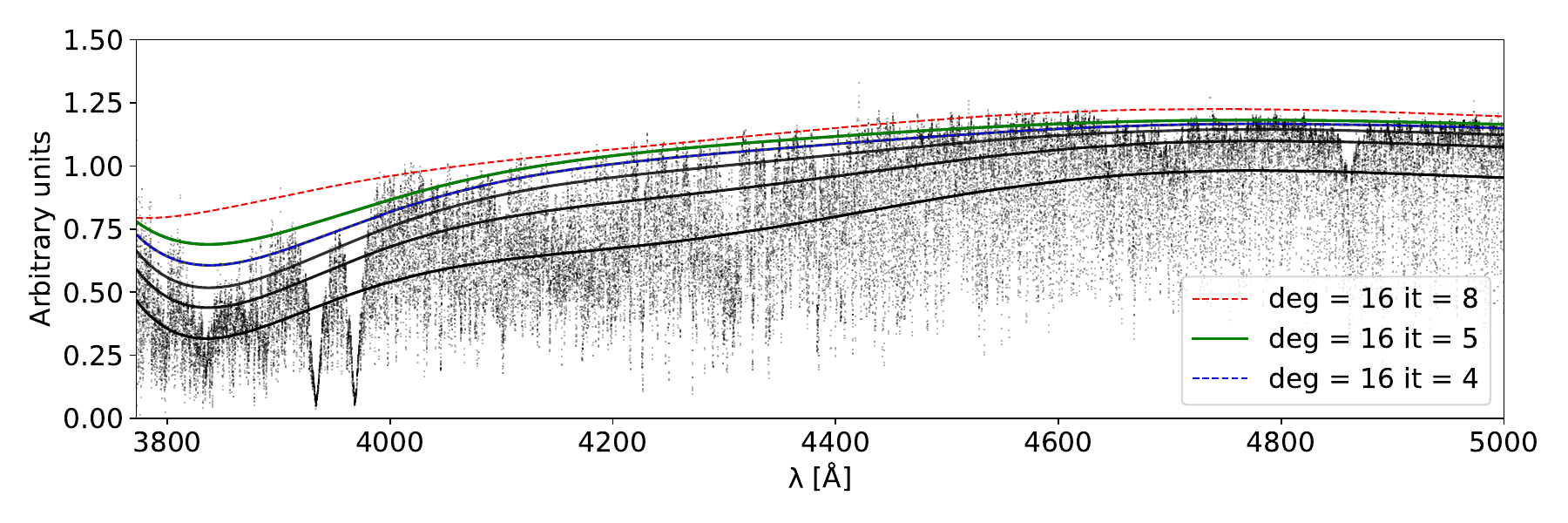}
    \caption{Illustration of the iteration process in a given Chebyshev polynomial order for an early-type star (top) and a late-type star (bottom), focussing on the wavelength range [3750, 5000]~\AA. Top: Order 26 for HD~3901 with spectral type B2V. Bottom: Order 16 for HD~41116 with spectral type G5III. The successive polynomial fits start with the thick black line, representing the first iteration, and progress towards the green line, representing the iteration with the best $d$ score. For the sake of completeness, the iterations corresponding to the best $p$ and $n$ scores are represented as well, in red and blue respectively. The final selection is based on $d$, combining $p$ and $n$.}
    \label{fig:poly_fit}
\end{figure*}

We first smoothed the result of the division of the calibrator spectrum by the model in Eq.~\ref{eqn:hatre} with a median filter and then fitted it with a spline in order to produce a smooth yet accurate representation of the effective response before injecting it in Eq.~\ref{eqn:Sfinal}. This was necessary in order to avoid introducing the noise resulting from the observation of the calibration star or local residuals from its model into the corrected spectra. The density of the knot points used to fit the spline was chosen locally, depending on the complexity of the response. Due to the intrinsic width of H$\alpha$, we noted that the spline occasionally overfitted that spectral line, introducing some local features at the few percent level around it in the response. Hence, we recommend some caution for this line, especially for nights where the response was derived from \object{GSC4293-0432} or \object{HD\,46300} (identifiable from the fits header entry {\sc stdname}; see Appendix~\ref{appendix}).

Once the effective response was established for the night, all spectra obtained over the corresponding night were divided by it, resulting in a reliable representation of the continuum shape. The derivation of the instrumental response for a night is illustrated in  Fig.~\ref{fig_respDerivation}, and an example of response-correction of a science-spectrum is displayed in Fig.~\ref{fig_FullCorrHD3379}.

At this stage, three of the coldest objects accepted as calibration stars in the observation planning had to be rejected because some low-frequency component of their continuum was not properly reproduced by the models, thus inducing systematic residual features in the instrument response model (HD\,220657, HD\,226826, and HD\,185395, with $T_\mathrm{eff}=6000$, 6500, and 6900~K, respectively). The corresponding nights were considered not to have any reference spectrum for the response derivation. 
Including those nights, no calibration star could be observed for over 129 of the 479 nights encompassed by this programme, corresponding to 1007 spectra in total (31\%). The effective response derived for the nearest night in time was then used instead. We find it important to note that the shape of the flat-field affects the effective response. This change in the $FF_{SED}$ depends on the temperature of the lamps, which is not monitored. Hence, we cannot control the applicability of a given response (derived from another night) to a given night. As a result, for such nights, most of the response corrections have a higher degree of inaccuracy, which may lead to some low-frequency signatures in the instrument-response corrected spectra. In order to easily identify the corresponding spectra, we introduced a dedicated Boolean keyword called {\sc `stdnight'} in the fits headers, and set it to zero when the response could not be computed from calibration observations obtained on the same night (see Appendix~\ref{appendix}). 

To quantify the uncertainty introduced by correcting spectra with a response established from another night, we first re-scaled all responses to an average of one over the range [5400, 5600]\,\AA. We then computed the ratio of each response by that of the next night in our programme. The standard deviation of the resulting dataset is presented in red in Fig.~\ref{fig_stddev_resp}. The additional uncertainty is limited to less than 5\% over the range between 4100 and 6750\,\AA. Beyond 7000\,\AA, the variability compared to our reference range is significantly larger, peaking at almost 17\% at 7950\,\AA. This is expected because, as stated above, that part of the spectrum is illuminated with a different lamp during the flat-field measurements, hence inducing larger differences between the bluest and reddest parts of the response as the two lamps may react differently to the various measurement conditions.

For the sake of completeness, we also analysed the variability of the response over the entire programme. Therefore, we computed the median response and divided all of the responses by it. We then computed the standard deviation of the result as a function of the wavelength. The result is displayed in Fig.~\ref{fig_stddev_resp} and indicated by the grey dashed line. Although the variability over the entire programme is significantly larger than between subsequent observing nights, hence suggesting some kind of monotonic evolution of the response, we observed a more irregular behaviour. Our efforts to identify the driving parameters for that variability (e.g. environmental parameters or ageing of the lamps) nevertheless remained unsuccessful.

\begin{figure*}[bht]
    \centering
    \includegraphics[width=0.49\linewidth]{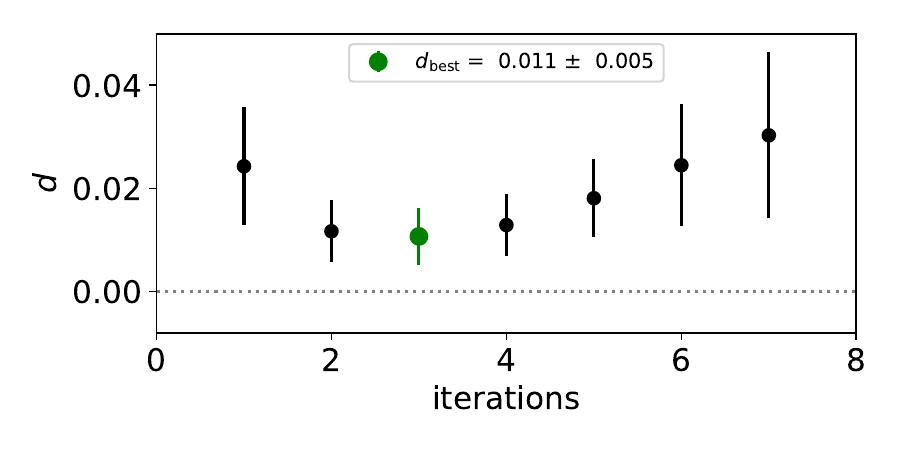}
    \includegraphics[width=0.49\linewidth]{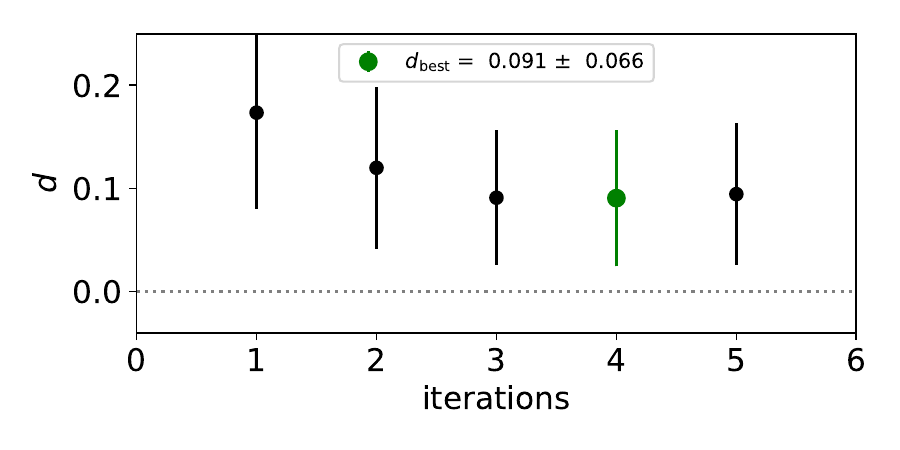}
    \caption{Evolution of the $d$ scores with the number of iterations for a given polynomial order applied to the same spectra as in Fig.~\ref{fig:poly_fit}. Left: Order 26 for HD~3901 with spectral type B2V Right: Order 16 for HD~41116 with spectral type G5III. 
    The numbers given below each point represent the number of the selected wavelength points (over the full wavelength range) in the polynomial fitting.}
    \label{fig:poly_iteration}
\end{figure*}

\begin{figure*}
    \centering
    \includegraphics[width=\linewidth]{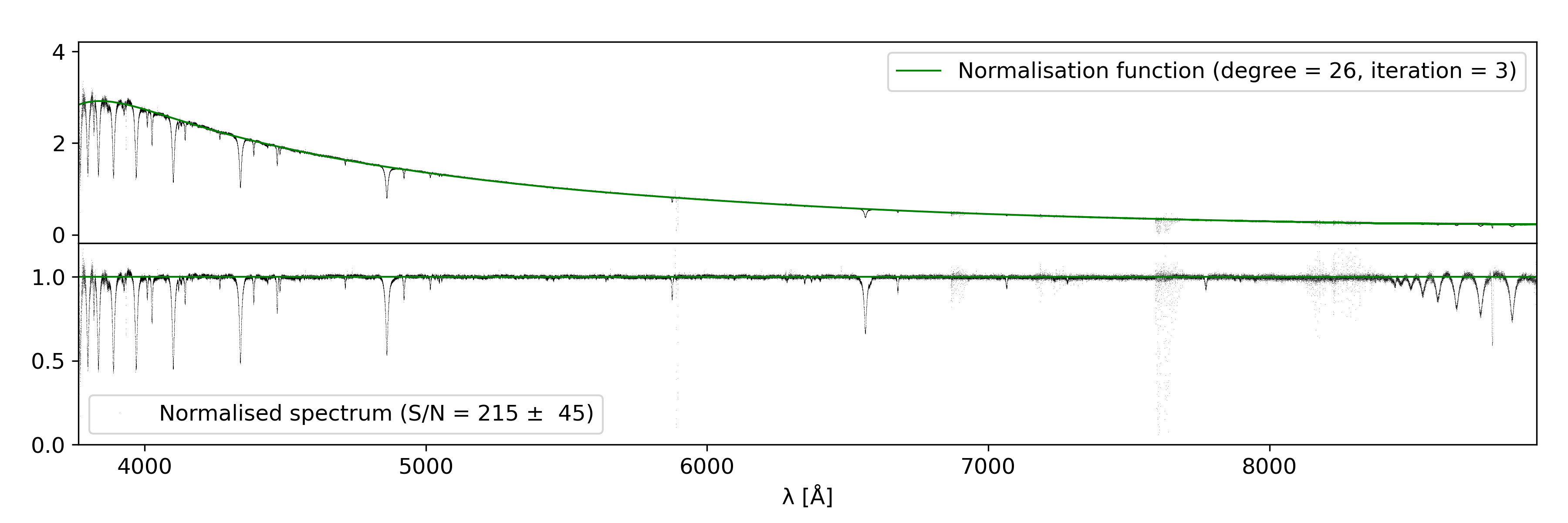}
    \includegraphics[width=\linewidth]{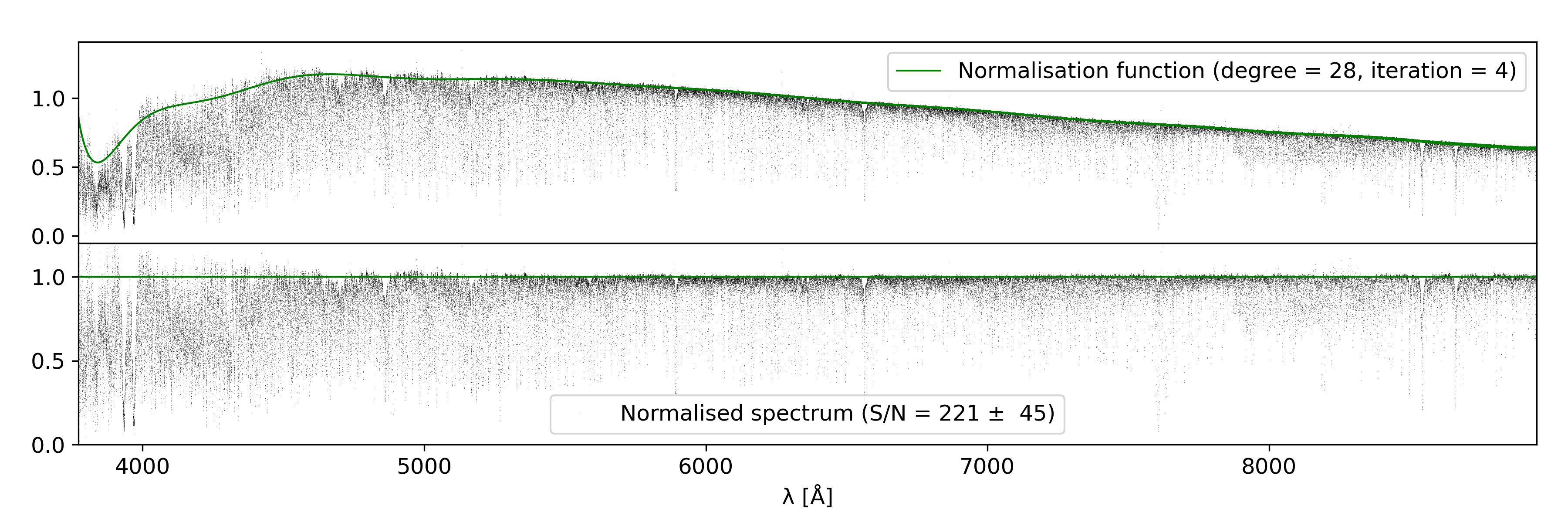}    
    \caption{Pseudo-normalised version of the observed spectra presented in Fig.~\ref{fig:poly_fit} but over the full wavelength range. Top: HD~3901 (B2V). Bottom: HD~41116 (G5III). Top sub-panel: Response-corrected spectrum (black dots) and the normalisation function (green line). Bottom sub-panel: Normalised spectrum.} 
    \label{fig:norm_spectra}
\end{figure*}

\subsection{Normalisation}
The procedure to pseudo-normalise the HERMES spectra was eased by the previous steps of the data reduction, but it remained challenging since the spectral range to normalise runs over more than 5\,000~\AA. Manual pseudo-normalisation tools exist, 
but they render the normalisation process highly time-consuming and do not ensure the repeatability of the result. Recent development to make automatic and tunable normalisation processes has emerged based on the convex-hull and alpha-shape theories, such as AFS  \citep{Xu2019} and  RASSINE \citep{Cretignier2020},  and even based on machine learning techniques, for example, using deep neural networks \citep{Rosanski22}. In this work, we rely on a more classical approach based on a iterative filtering method similar to the ones that use a sigma-clipping process, as done, for example, on a smaller wavelength extent in \cite{Bermejo2013,Sana2013,Paletou15}. Given that our approach to normalisation relies on the pseudo-continuum as measured rather than on the true continuum of the object, it underestimates the continuum in the blue, especially for cold objects, and we hence refer to it as `pseudo-normalisation'.

It is often argued that high-order polynomials are not flexible enough. This is true when the normalisation process occurs directly on the merged \'echelle spectra (top panel of Fig.~\ref{fig_FullCorrHD3379}), which is not corrected from the instrumental response function (middle panel of Fig.~\ref{fig_FullCorrHD3379}). In our case, the polynomials are fitted to the response-corrected spectra, whose shapes are much smoother (bottom panel of Fig.~\ref{fig_FullCorrHD3379}), except for late M-type stars, where strong molecular bands render this approach inappropriate (and the normalisation is skipped; see below).

Our normalisation process can be divided into five steps: First, we smooth the response-corrected spectrum using a Savitzky–Golay filter with a third-order polynomial and a window length of 5 pixels. This allows for the smoothing of any possibly remaining cosmics or noisy parts of the spectrum (mainly below 4\,000~\AA). Second, for computational reasons, we undersample the smoothed version of the response-corrected spectrum by a factor 20, resulting in spectra with about 8\,400 samples evenly spaced in log-wavelength. Third, we compute a grid of normalising functions. Those functions are Chebyshev polynomials of different degrees between zero and 40. Each function results from an iterative process described in the next paragraph. Fourth, a score is attributed to each of the normalisation functions reflecting their quality. Fifth, the score is used to select the best normalisation function in the grid.

The iterative process in the third step goes as follows: After the polynomial fit, we filter out all samples from the spectrum falling under the fit and proceed to fit the same polynomial on the remaining samples. At each iteration, a normalised spectrum is computed whose quality is assessed by a score `$d$':
\begin{equation}
    d = p - n, 
    \label{eq:dscore}
\end{equation} 
where $p$ is the median of all flux samples above one and $n$ is the median of all samples below one. As the best estimate of the normalised spectrum, we accept the result realising the smallest score $d$ over all iterations and all polynomial orders ($d$ is positive by construction).

This process nominally runs over ten iterations but is occasionally halted before by an additional stop-criterion that imposes a minimum sampling density over the entire wavelength range. We indeed imposed that at least 100 wavelength samples remain in the filtered spectrum per range of 1000~\AA. Without this, the filtering, which is increasingly aggressive as iterations progress, may result in gaps in the wavelength coverage, leading to ill-behaved solutions to the fits.

An example of the selection of the best function for a given polynomial order is illustrated in Fig.~\ref{fig:poly_fit} for an early-type (top panel) and a late-type spectrum (bottom panel). Fig.~\ref{fig:poly_iteration} shows the behaviour of the $d$ score computed in the course of the fitting process of the same spectra, that is, the successive iterations for a given polynomial order.

The $d$ score is closer to zero when fewer absorption lines are present in the spectra (compare the left and right panels of Fig.~\ref{fig:poly_iteration} for a B2V-type star with fewer absorption features than the G5III-type star). High $d$ scores reflect a poorer normalisation but may also be correlated with the decreasing sampling in wavelengths due to the filtering, as explained above. Of course, the error bars on the $d$ score are also an indication of the quality of the normalising functions. The error bars on $d$ are based on the median absolute deviation for scores $p$ and $n$, as this is more robust to outliers than the standard deviation, and propagated to $d$ via Eq.~\ref{eq:dscore}.


Figure~\ref{fig:norm_spectra} illustrates the final normalising functions and normalised spectra for the stars presented in Fig.~\ref{fig:poly_fit}. A polynomial degree 26 at the third iteration gives the best $d$ score for HD3901 (B2V), and a polynomial degree 28 at the fourth iteration gives the best $d$ score for HD~41116 (G5III).

\begin{figure}
    \centering
    \includegraphics[width=\linewidth]{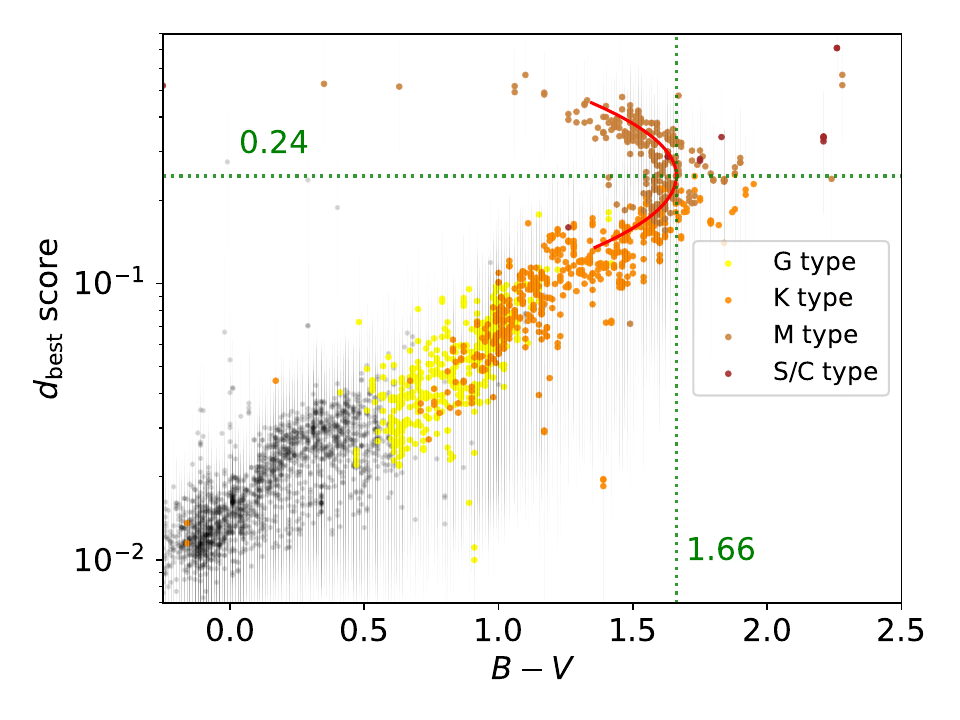}
    \caption{Best $d$ score as a function of $B-V$. We provide normalised spectra when the best $d$ score is lower than 0.24. The spectra excluded from the normalisation process represent about 6\% of the sample and are made of M-, S-, and C-type stars. See text for details.}   \label{fig:dbest_BV}
\end{figure}

The presence of telluric lines at the red end of the spectra resulted in poor access to and hence a poor representation of the continuum longwards of 8950\,\AA. This lead us to discard the last 50~\AA\ of the normalised spectra.

Finally, we decided not to provide normalised spectra for the latest spectral types presenting significant molecular bands. We used a criterion based on the $d$ score for the cut-off. The score indeed shows a positive trend with $B-V$, as displayed in Fig.~\ref{fig:dbest_BV}. A turn-off appears around $B-V=1.6$. It is produced by long-period variables, which are all (asymptotic) red giants of spectral types M to C \citep{lebzelter2023}. For the giants (luminosity class III), which make up the bulk of our sample of M-type stars, the $B-V$ color index reaches a maximum value of 1.65 for M5, then decreases to 1.50 for M8, as shown in Table~1 of \citet{Fitzgerald1970}. The main sequence does not show this turnover, but our sample only includes three main-sequence M-type stars, explaining the depletion at larger $B-V$. Hence, to define the cut-off for the normalisation, we adjusted a quadratic $B-V$ function with respect to the $\log{d_\mathrm{best}}$ score. This provided a maximum value of the $d_\mathrm{best}=0.24$ for which we provide normalised versions of the spectra. We confirmed that the criterion is appropriate by visual inspection. The median $B-V$ of the spectra with larger $d_\mathrm{best}$ is 1.6. The excluded spectra represent about 6\% of the MELCHIORS database (201 spectra corresponding to 156 stars) and are mainly made of M giant and sub-giant stars as well as S- and C-type stars, with significant expression of molecular bands. 

\section{Results}
\label{sec:results}

\begin{figure*}
\centering
\includegraphics[width=\hsize]{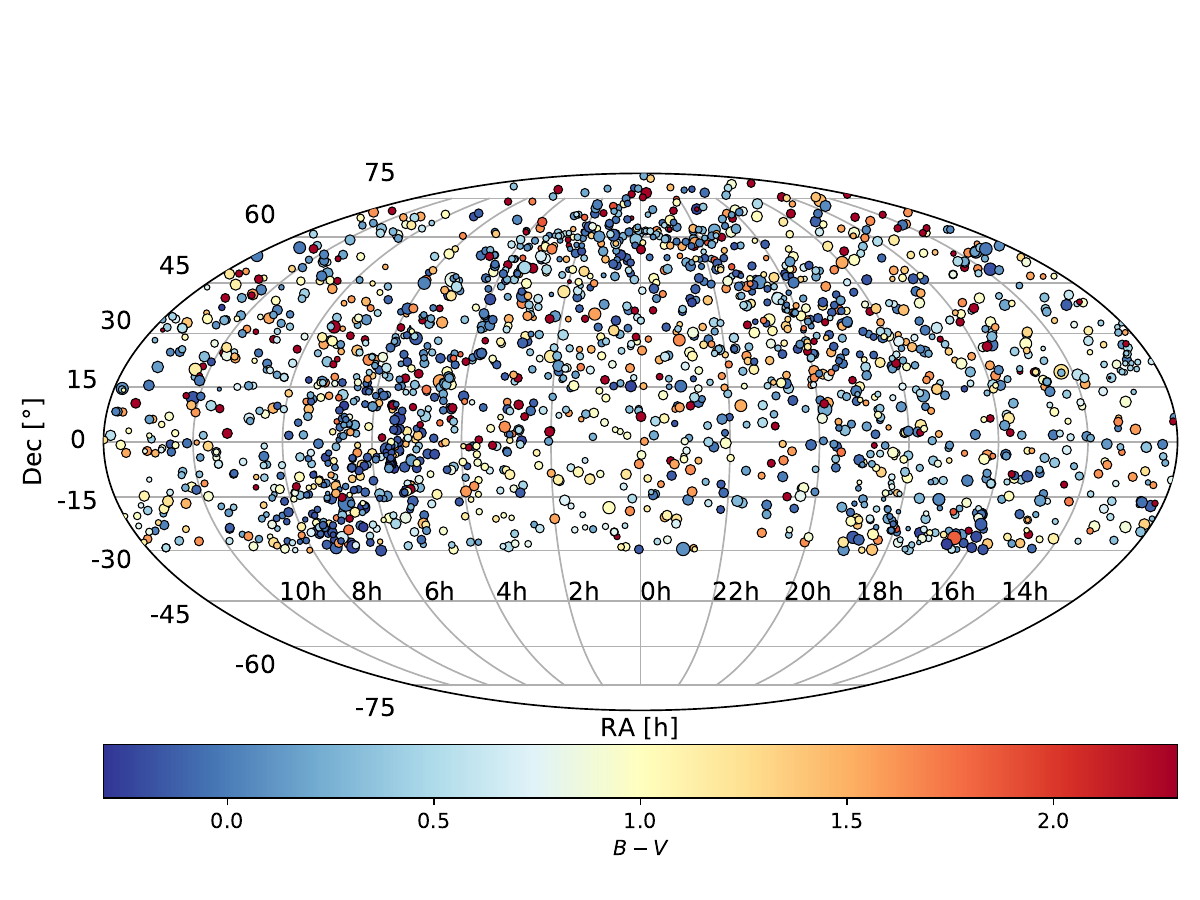}
     \caption{Sky distribution of the objects from the MELCHIORS catalogue. The size of the symbols is related to their apparent V magnitude.}
     \label{fig:sky_distribution}
\end{figure*}

\begin{figure}
\includegraphics[width=\linewidth]{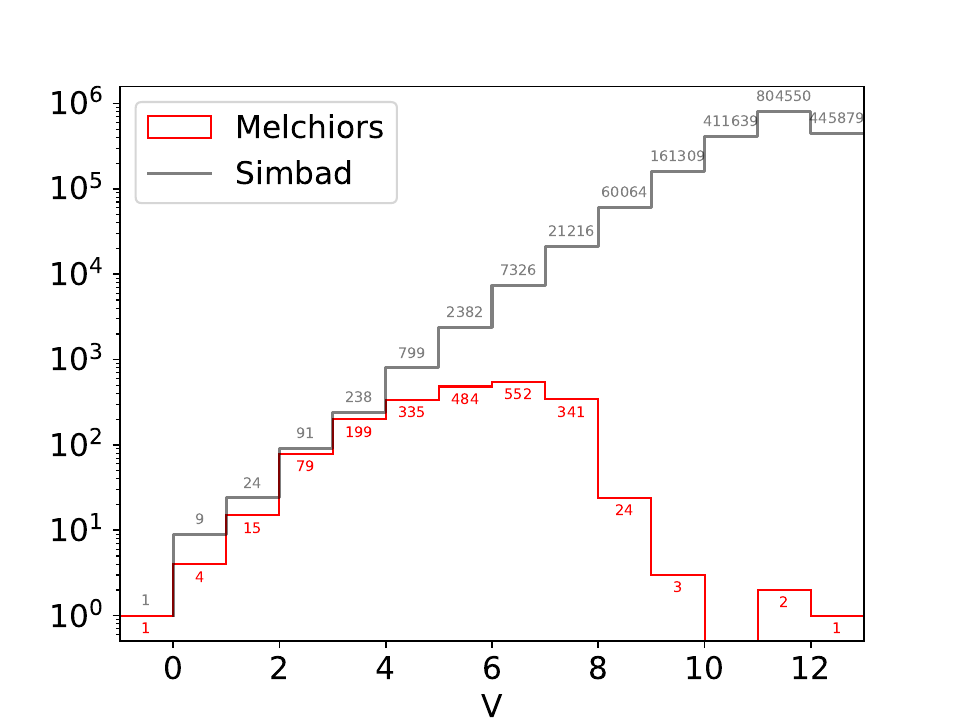}
\caption{ V-band magnitude distribution of the MELCHIORS stars compared to the stars listed in Simbad with declinations larger than $-30^\circ$. The median $V$-band magnitude of MELCHIORS stars is 5.9.}
\label{fig:vmag_histo}
\end{figure}

\begin{figure}
\centering
\includegraphics[width=\linewidth]{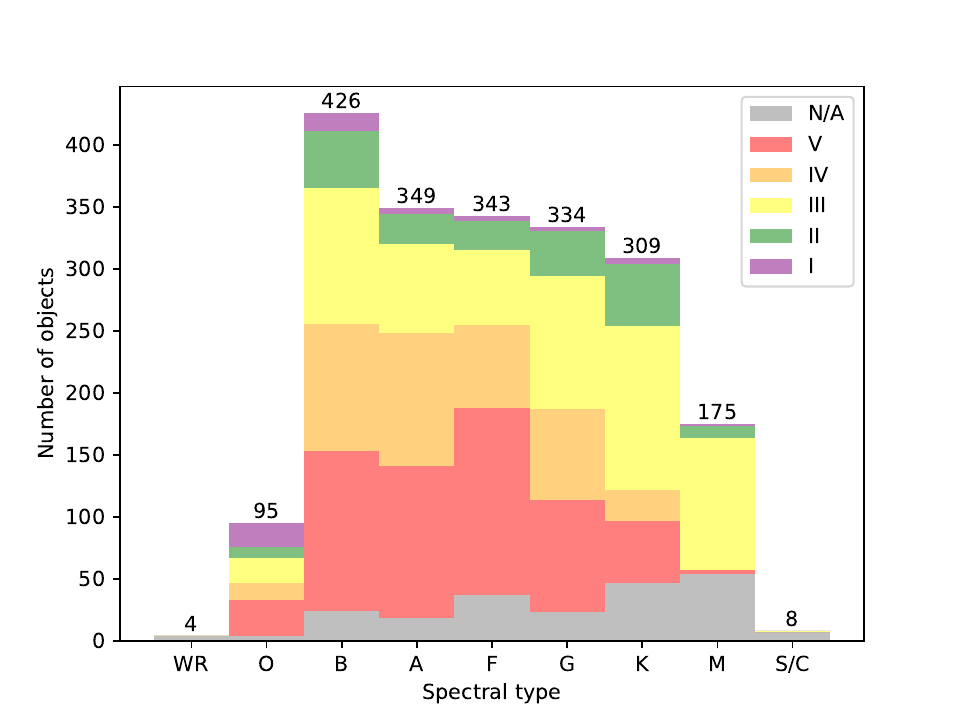}
\caption{Distribution of the MELCHIORS stars per spectral type and luminosity class (excluding the Sun, as in all other subsequent figures). The number above each bar is the total number of star per spectral type. About 11\% of targets do not have a luminosity class on Simbad (grey histogram). Early-type stars and giants are over-represented in the database.}  
\label{fig:sptype_hist}
\end{figure}

\begin{table*}
\centering
\caption{Number of stars per spectral type and luminosity class (excluding the Sun).}
\begin{tabular}{rrrrrrrrrrr}
\hline\hline
      &   WR &   O &   B &   A &   F &   G &   K &   M &   S/C & Total \\
\hline
\\
V     &    0 &  29 & 129 & 123 & 151 &  90 &  50 &   3 &    0  & 575\\
IV    &    0 &  14 & 103 & 107 &  67 &  73 &  25 &   0 &    0  & 389\\
III   &    0 &  20 & 109 &  72 &  60 & 107 & 132 & 107 &    1  & 608\\
II    &    0 &   9 &  46 &  24 &  24 &  37 &  50 &   9 &    0  & 199\\
I     &    0 &  19 &  15 &   5 &   4 &   3 &   5 &   2 &    0  &  53\\
N/A   &    4 &   4 &  24 &  18 &  37 &  23 &  47 &  54 &    7  & 218\\  
\\
\hline
Total &    4 &  95 & 426 & 349 & 343 & 333 & 309 & 175 &    8  & 2042\\
\hline
\end{tabular}
\label{tab:stats}
\end{table*}

\subsection{Stellar content}

The MELCHIORS database is made of 2043 stars (excluding the Sun) all above a declination of $\delta =-30^\circ$ since they were observed from the island of La~Palma. Figure~\ref{fig:sky_distribution} shows the distribution of the MELCHIORS targets on the sky (in equatorial coordinates). 
An overdensity of early-type stars traces the disc of the Milky Way.
The targets are colour-coded with their colour index $B-V$. The lowest values, representing the hottest stars, mainly trace the disc and bulge, while the highest values, representing the coolest, reddest, and the most likely to be giant stars, are rather located in the halo. The distribution of the targets in the $V$-band magnitude is displayed in Fig.~\ref{fig:vmag_histo} and includes the brightest stars of the northern hemisphere. The median $V$ magnitude of the targets is $5.9$. We compared our sample with the absolute numbers of stars in the Simbad database with a declination larger than $-30^\circ$, in bins of one magnitude.\footnote{These numbers were obtained by using the `Simbad: query by criteria' with the search expression \texttt{Vmag >= X \& Vmag < X + 1  \& dec > -30 \& otypes = '*'}, where \texttt{X} varies between $-1$ and $12$ per step of 1 magnitude. The latest condition on \texttt{otypes} ensures the exclusion of the non-stellar objects from the query.} The completeness of the MELCHIORS database compared to Simbad is  $\sim80$\% for $V\le4$ but drops to 55\% for $V\le6$. Excluding the Sun, the apparent brightest target is Arcturus ($\alpha$~Boo), while the faintest star is \object{HD~286340,} which is an SC7/8 star. 

Most of the MELCHIORS stars have spectral types and $V$-band magnitudes available in Simbad.
We cross-matched targets without a spectral type or $V$-band magnitude with the catalogue of Stellar Spectral Classifications \citep{Skiff14} in order to gather the missing information. For \object{HD~88133}, we adopted the Gaia magnitude because the information was not available in \cite{Skiff14}. For \object{HD~114762}, we adopted the V magnitude of the primary object ($V=7.3$; removed from Simbad in 2018) rather than the most recent reference in~\cite{Skiff14}, corresponding to the very faint M9 companion ~\citep[$V=15$;][]{patience2002,bowler2009}.

Figure~\ref{fig:sptype_hist} presents the number of MELCHIORS stars per spectral type (OBAFGKM) also including a few Wolf-Rayet (WR) and evolved S-type and C-type (S/C) stars. Due to the bias of our sample towards low apparent magnitudes, this distribution does not reflect the stellar initial mass function. Instead, it is biased towards the brightest and hottest stars, while about 99\% of all stars are of a spectral type later than A. The luminosity classes are also biased. While we expected about 10\% of the stars to have evolved off the main sequence, giants and supergiants  (luminosity classes I, II, and III) represent between one-quarter to more than one-half of the number of stars in our sample, as a function of the spectral type. About 11\% of the targets do not have a luminosity class in the Simbad database and are represented in grey in Fig.~\ref{fig:sptype_hist}. The summary of the MELCHIORS database per spectral type and luminosity class is given in Table~\ref{tab:stats}.

\begin{figure}
\includegraphics[width=\linewidth]{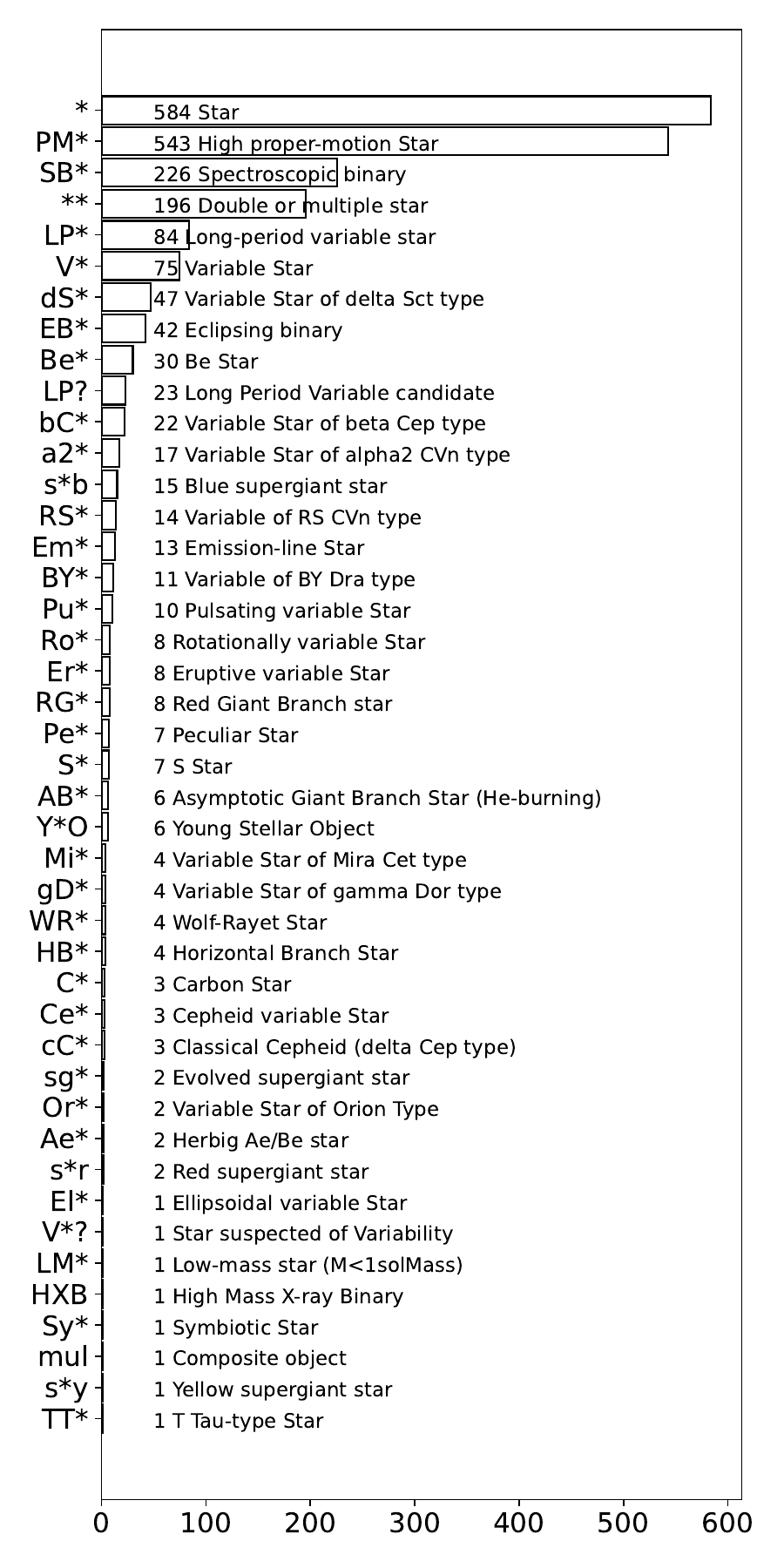}
\caption{ Simbad types of stars in the MELCHIORS database.}
\label{fig:simbad_type}
\end{figure}

 We display the distribution of MELCHIORS stars per Simbad type in Fig.~\ref{fig:simbad_type}. About 29\% of the objects are normal single stars, and 27\% are flagged as high proper motion stars. We stress that about 24\% of the library is related to stellar multiplicity (227 spectroscopic binaries, 196 double or multiple stars, 42 eclipsing binaries, 22 variable star of $\beta$~Cep type, 1 high-mass X-ray binary, 1 symbiotic star, 1 composite object, and 1 ellipsoidal variable star). This gives a general overview of the type of objects in the MELCHIORS database as reported in Simbad. (Nonetheless, we advise the reader to keep in mind that the Simbad database is constantly being updated and that the distribution of object types reflected in Fig.~\ref{fig:simbad_type} provides but a snapshot of the knowledge gathered in the Simbad database around March 2022.)

\begin{figure*}
    \centering
    \includegraphics[width=1.1\linewidth]{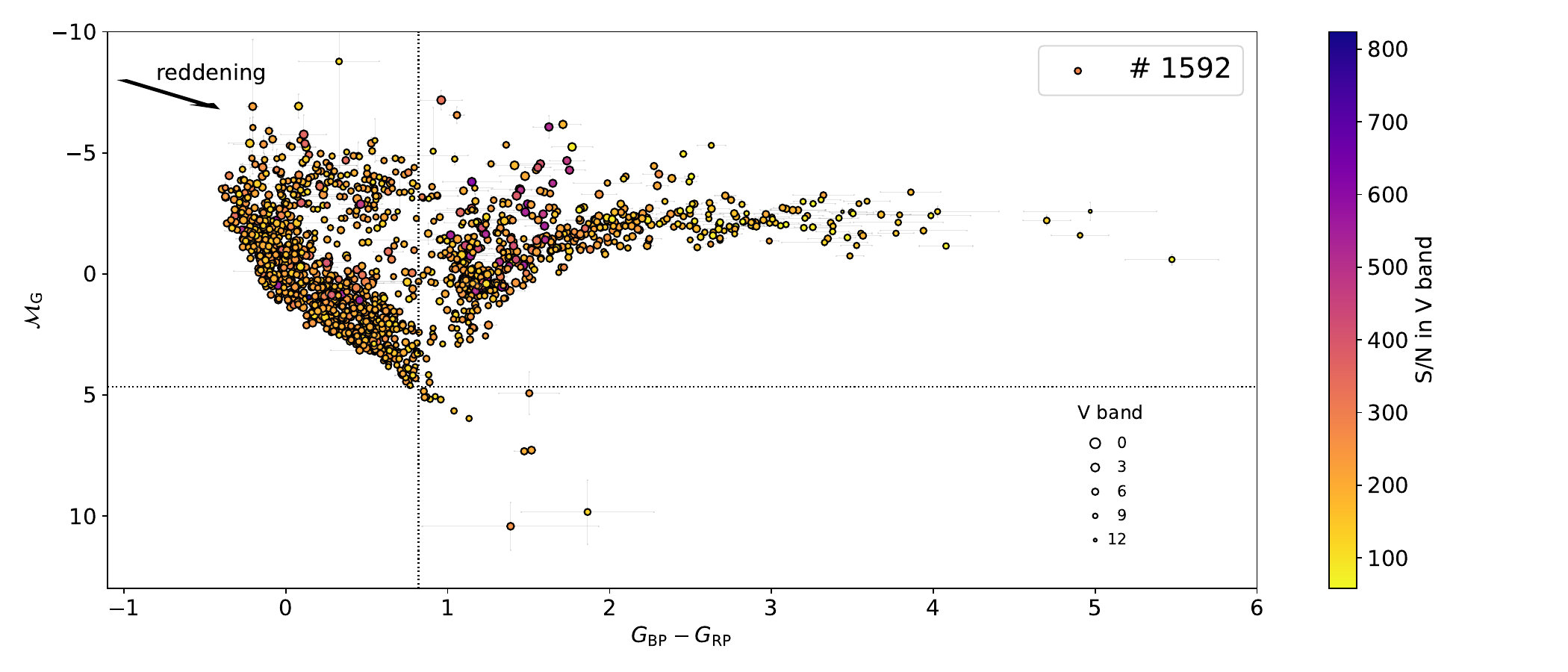}
    \caption{Colour-absolute magnitude diagram of the MELCHIORS stars using astrometric and photometric Gaia DR3 data. Apparent visual magnitudes and S/N of the best spectrum per star are also encoded. The intersection of the dotted lines shows the position of the Sun.}
    \label{fig:camd}
\end{figure*}

\begin{figure}
    \centering
    \includegraphics[width=\linewidth]{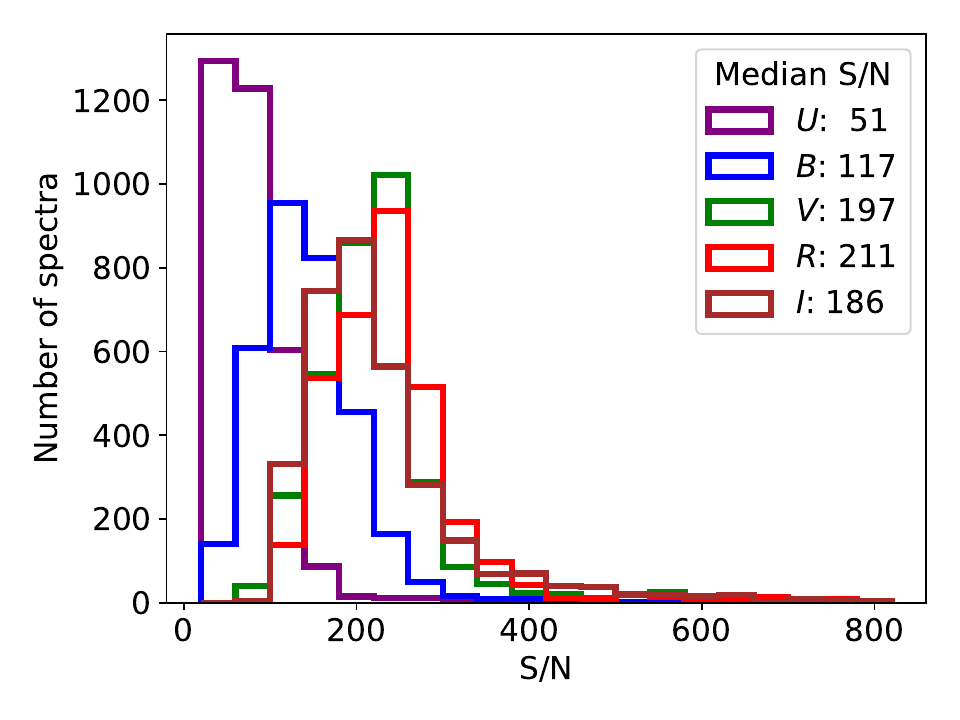}
    \caption{Distribution of the spectra as a function of S/N in the U, B, V, R, and I wavelength ranges.}
    \label{fig:snr_histo}
\end{figure}

\begin{figure}
    \centering
    \includegraphics[width=\linewidth]{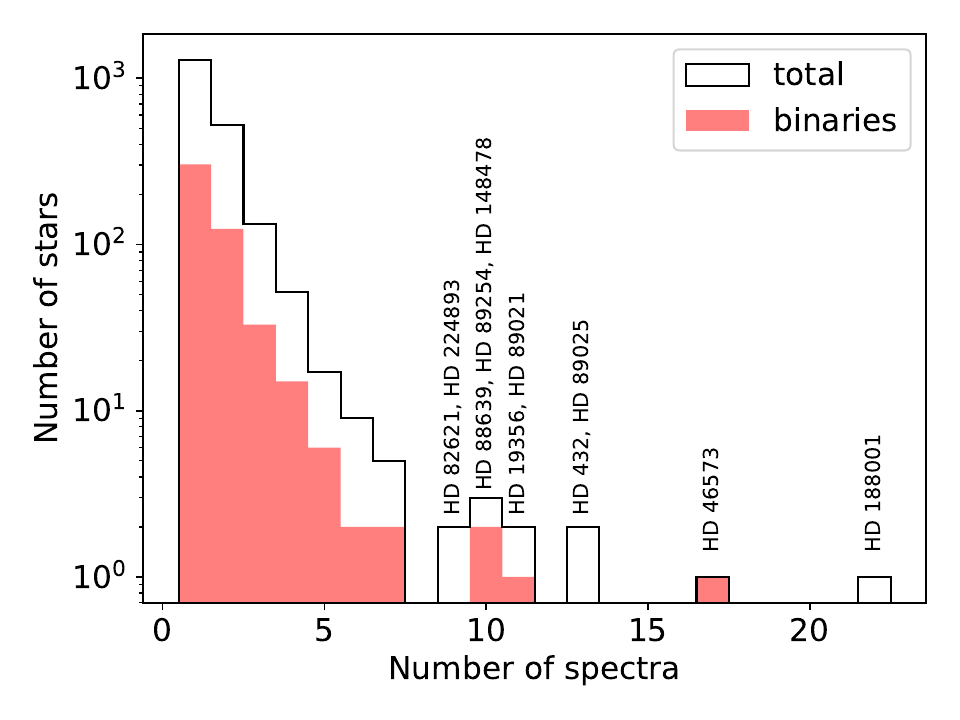}
    \caption{Number of spectra per star on a logarithmic scale. The distribution of known Simbad binaries is overplotted in red. The name of the stars with more than eight spectra are indicated. The MELCHIORS star with the highest number of spectra, HD188001 (9 Sge),is an O-type runaway blue supergiant, that could also be a spectroscopic binary (see text for details).}
    \label{fig:spectra_star}
\end{figure}

\begin{figure*}
 \centering 
 \includegraphics[width=\linewidth]{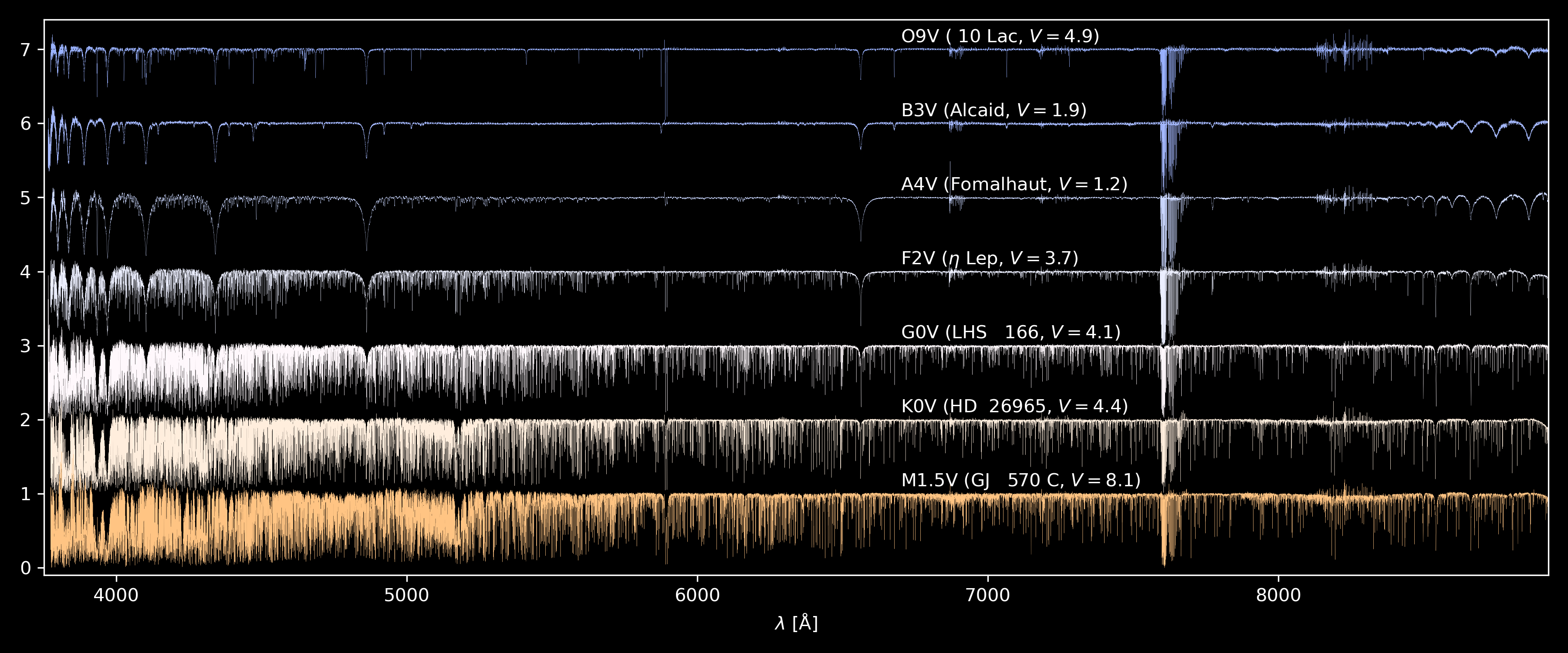}
 \includegraphics[width=\linewidth]{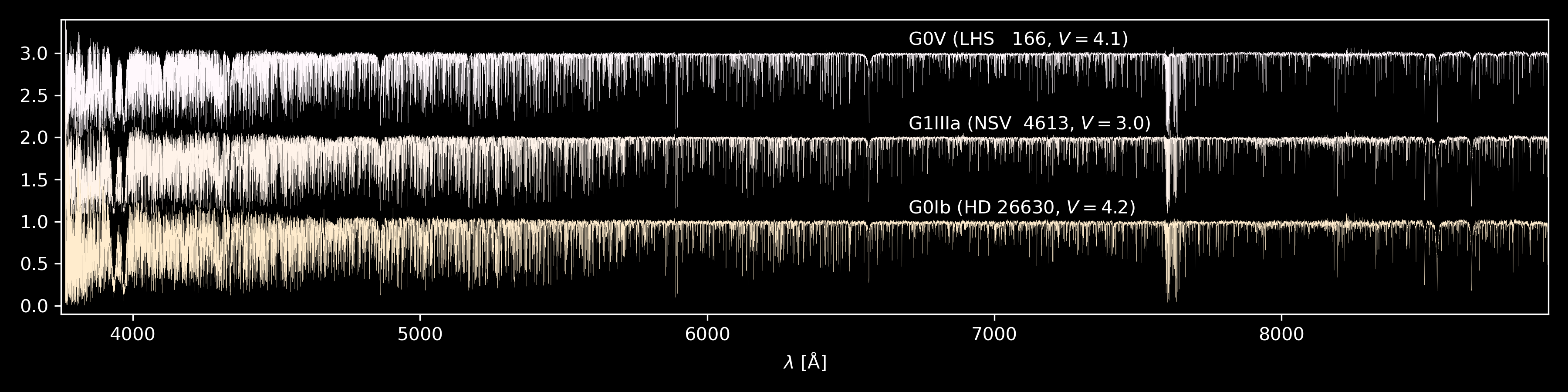}
\caption{Examples of normalised spectra. Top: Spectral type sequence -- OBAFGKM. Bottom: Luminosity class sequence -- V, III, I. We use a black background for a better visualisation of the colours of the corresponding stars, following \href{http://www.vendian.org/mncharity/dir3/starcolor/UnstableURLs/starcolors.html}{M. Charity's work}, who derived chromaticity and rgb-pixel colour from spectra for various stellar types and/or classes, providing physically motivated colours.}
\label{fig:nspectra}
\end{figure*}

\subsection{Gaia colour-absolute magnitude diagram}
We cross-matched the MELCHIORS database with Gaia DR3 using a cone search radius of 1 arcsec that was iteratively increased if no match was obtained until 3 arcsec at maximum. We obtained a positive cross-match for a sub-sample of 1690 of the 2043 stars. Among them, one has a negative parallax, and 97 do not have any parallax at all. These mainly correspond to very bright targets.
A colour-absolute magnitude diagram of the MELCHIORS stars is displayed on Fig.~\ref{fig:camd} for the sub-sample of 1592 stars for which parallaxes and photometry are available from Gaia DR3. The main sequence is poorly populated in G, K, and M stars compared to the giant branch for those types, due to a combination of two factors. First, the magnitude-limited selection imposes a bias against cool main-sequence stars with respect to their giant counterparts
(see Fig.~\ref{fig:sptype_hist} and Table~\ref{tab:stats}). Second, the fraction of high proper-motion stars is important (80 of the $\sim$140 G, K, and M main-sequence stars), and those stars were filtered out by the cone search radius. The error bars on the colour and G magnitude were computed according to Equation~12 of \cite{Merle20} since it is not directly provided in the Gaia DR3. The uncertainties on the absolute magnitudes were obtained by propagating errors on parallaxes and G magnitudes. The position of the Sun is represented with the dotted line at $G_\mathrm{BP} - G_\mathrm{RB}=0.82$ and $\mathcal{M}_G = 4.67$ \citep{Casagrande18}. The asymptotic giant branch (AGB) reaches a saturation in absolute $G$ magnitude due to strong molecular absorptions in this wavelength range, explaining the horizontal shape of the AGB in the Gaia colour-absolute magnitude diagram.

\subsection{The spectra}

The MELCHIORS database is made of 3256 high S/N spectra of the 2043 selected stars covering the wavelength range [3800, 9000]~\AA\ with the spectral resolution of $R=85\,000$ of the HERMES spectrograph on the Mercator telescope. Figure~\ref{fig:snr_histo} shows the S/N distribution of the spectra at the effective wavelengths of the U, B, V, R, and I filters, as estimated from the counts in the corresponding wavelength bins of the original 2D spectra (spot checks in the final products indicate this to be conservative). Except in the U band, the median S/N is larger than 110 in all bands, and it reaches 200 in the V and R bands.

Figure~\ref{fig:spectra_star} represents the number of spectra per star, and shows that 50\% of the stars have more than one spectrum. The binaries, overplotted in red in the figure, were selected from the main Simbad object types. They represent about one-fourth of the database. The stars with more than eight spectra are individually identified in the figure. 

We rejected HD\,205767 ($\xi$ Aqr) from the catalogue. Though it was seemingly acquired at the correct celestial coordinates, our spectrum did not match the spectral type of that object. Also, we did not investigate the stellar binary content of the MELCHIORS database in detail but relied on the main Simbad type. Consequently, we could have missed some binaries, or we may provide a simple view on debated objects. As an example, HD~188001 (9 Sge), a runaway O-type star, is the star with the highest number of spectra (23) in the MELCHIORS database. Its main object type in Simbad is blue supergiant, while it is also known as a spectroscopic binary in the SB9 database \citep{Pourbaix04}, with orbital parameters from \citet{Aslanov92}. Nevertheless, this star could also be a single star, according to \cite{McSwain07}, while \mbox{\citet{Aldoretta15}}, suggesting constant RV, mentioned the star as having an unresolved companion. 
We also mention that there are 56 non-single stars identified from the Gaia-DR3 cross-match: 24 astrometric binaries, 24 spectroscopic binaries, 7 astro+spectro binaries, and one eclipsing binary (the blue supergiant HD 167971). They can be retrieved using the meta-data described in Appendix~\ref{appendix:gaia}.
We note that an individual analysis of every object is nevertheless beyond the scope of the present paper. Finally, we illustrate the variety of spectra in our database by presenting a sample of spectra over a sequence of spectral types (OBAFGKM) and luminosity classes (V, III, I) in Fig.~\ref{fig:nspectra}. 

\begin{figure*}
\includegraphics[width=\linewidth]{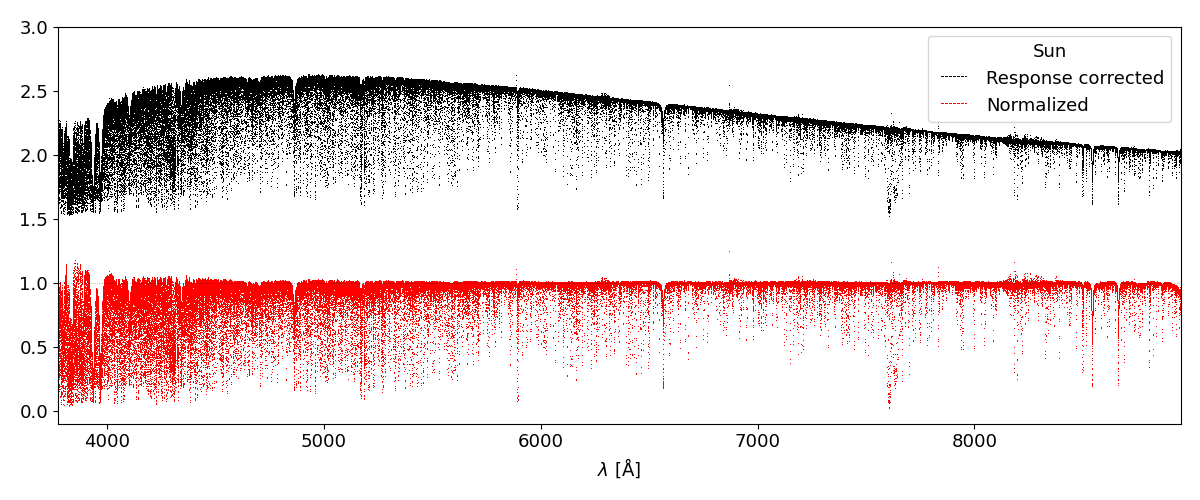}
\includegraphics[width=\linewidth]{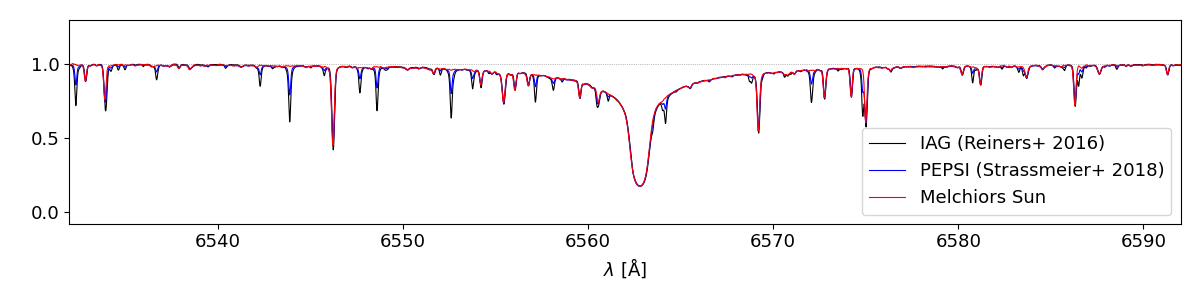}
\caption{MELCHIORS solar reference. Top: MELCHIORS solar spectrum. The response and telluric corrected spectrum is shown in black and the normalised spectrum is shown in red. Bottom: Excerpt around H$\alpha$ compared to the IAG \citep{reiners2016} and PEPSI \citep{strassmeier2018} solar spectra degraded to the resolution of HERMES. Note the correction for the telluric lines performed in the MELCHIORS solar spectrum, in contrast with the other spectra.}
\label{fig:solar_spectra}
\end{figure*}

\subsection{The MELCHIORS solar reference}
The provided MELCHIORS solar spectrum reference was taken as a reflection spectrum on Ganymede on June 22, 2018. It has been reduced using the same procedure as all the other spectra (see Sect.~\ref{sec:data_proc}). The spectrum is set in the barycentric restframe of the Solar System and corrected from the radial velocity of Ganymede. The S/N is about 560 at 5000~\AA. Figure~\ref{fig:solar_spectra} presents the solar spectrum corrected from instrumental response and from tellurics in the top panel. As already mentioned for the other stars, we noticed an imperfect normalisation at both ends of the spectrum. In the blue, this is due to the presence of many metallic lines, including strong ones (the \ion{Ca}{II} H and K lines at 3968.47 and 3933.66~\AA, the \ion{Fe}{I} L line at 3820.44~\AA, and others, including strong lines of  \ion{Mg}{I}). In the red, this is due to the presence of telluric lines. In the bottom panel of Fig. \ref{fig:solar_spectra}, we show a focus around H$\alpha$ at 6562.81~\AA. We compared it with two recent observed solar spectra: (i) the IAG solar flux atlas from a Fourier transform spectrograph with a resolution of $R = \lambda/\Delta\lambda \sim 10^6$ \citep{reiners2016} and (ii) the PEPSI solar spectrum at the LBT with a resolution of $R=270\,000$ \citep{strassmeier2018}. The two comparison spectra were degraded at the resolution of HERMES by convolving them with a Gaussian profile of width $\sigma=3.5$~km/s, which corresponds to a resolution of $R=85\,000$. The comparison shows that the radial velocity correction was well performed. We note that the MELCHIORS reference spectrum is well corrected for the tellurics, which are clearly visible in the two other spectra at different intensity levels.  Our solar reference spectrum also shows excellent agreement with the solar atlas from \citet{Beck2016}, also produced with HERMES, as well as with the high-resolution spectrum from \citet{Wallace2011}. Interestingly, and with the exception of the ELODIE \citep{Prugniel07} and Gaia-FGK libraries \mbox{\citep{Blanco14}}, no solar spectrum is provided in the optical spectral libraries reported in Table~\ref{tab_Libraries}. 

\section{Summary}
\label{sec:conclusions}

The spectral library presented in this work gathers 3256 spectra of 2043 stars of all spectral types and luminosity classes, including the Sun as a star (with a spectrum of Ganymede). The spectra cover the full visible spectral range [3\,800, 9\,000]~\AA\ with high resolution {\bf ($R=85\,000$)} and with a median S/N reaching 200 in the V band. 
In addition, the spectra were acquired in a consistent way and with a single instrument. This is, to our knowledge, a unique combination in comparison to existing optical empirical libraries.

Special care was taken in the data reduction, allowing us to provide the spectra in four different flavours: raw, corrected from the instrumental response, with and without correction from the atmospheric molecular absorption, and normalised (including the telluric correction). As with any automatic processing, we recommend using the final products with caution, especially for the analysis of wide spectral features such as H$\alpha$.

The sample is $\sim 80\%$ complete up to $V=4$ for declinations $\delta\,>-30^{\circ}$ and has a median magnitude of $V=5.9$. The spectral library contains two or more spectra for about 50\% of the targets. All spectra are available in CDS as well as via an online repository: \href{https://www.royer.se/melchiors.html}{www.royer.se/melchiors.html}.

\begin{acknowledgements}
This work is based on observations made with the Mercator Telescope, operated on the island of La Palma by the Flemmish Community, at the Spanish Observatorio del Roque de los Muchachos of the Instituto de Astrof\'isica de Canarias. Mercator operates thanks to the IRI-FWO project with code I000521N. This work is also based on observations obtained with the HERMES spectrograph, which is supported by the Research Foundation of Flanders (FWO), Belgium, the Research Council of KU\,Leuven, Belgium, the Fonds National de la Recherche Scientifique (FNRS), Belgium, the Royal Observatory of Belgium, the Observatoire de Gen\`eve, Switzerland and the Th\"uringer Landessternwarte Tautenburg, Germany. The research for the present results has been subsidised by the Belgian Federal Science policy Office under contract No. BR/143/A2/BRASS. It has also received funding from the KU~Leuven Research Council (grant C16/18/005: PARADISE), from the Research Foundation Flanders (FWO) under grant agreement G089422N as well as from the BELgian federal Science Policy Office (BELSPO) through PRODEX grant PLATO. TM is supported by a grant from the Fondation ULB and by the BELSPO Belgian federal research programme FED-tWIN under the research profile Prf-2020-033\_BISTRO. JV acknowledges support from the Grant Agency of the Czech Republic (GA\v{C}R 22-34467S). MVdS acknowledges support from the European Union's Horizon 2020 research and innovation programme under the Marie Sk\l{}odowska-Curie grant agreement No 882991. M.K. acknowledges the support from ESA-PRODEX PEA4000127913. RK acknowledges the support by Inter-transfer grant no LTT-20015. PGB acknowledges support by the Spanish Ministry of Science and Innovation with the \textit{Ram{\'o}n\,y\,Cajal} fellowship number RYC-2021-033137-I and the number MRR4032204. SG acknowledges support from FWO by means of a PhD Aspirant mandate under contract No. 11E5620N.
\end{acknowledgements}

%
%

\bibliographystyle{aa}
\bibliography{p28_merged_tab}

\begin{appendix}
\section{Products and formats}
\label{appendix}

\begin{samepage}
The MELCHIORS database is accessible in CDS and can be found at \href{https:www.royer.se/melchiors.html}{www.royer.se/melchiors.html}, where simple selection tools and quick-look images are available as well. The library contains the following datasets:
\begin{enumerate}
    \item The spectral library itself
    \item The calibration spectra used to derive the instrumental response for every observing night
    \item The models of the calibration spectra
    \item The meta-data information, including\ 
    \begin{itemize}
        \item Parameters describing the spectra in the library and the corresponding targets
        \item Parameters linking the spectra with their associated calibration counterpart
        \item Parameters issued from the cross-match with Gaia-DR3, where applicable
    \end{itemize}
    \item Quick-look images
\end{enumerate}

\subsection{The spectral library}
\label{sec:library}
The spectra are presented individually and in FITS format, and they are identified by a unique observation identifier (hereafter `obsid'), which is a six-character integer occasionally pre-pended with `00'. The fits files have a primary header and one extension, which is a binary table with ten columns:
\begin{enumerate}
\item {\it wave\_log:} Natural logarithm of the wavelength
\item {\it wave:} Wavelength
\item {\it flux\_norm:} Normalised flux density
\item {\it err\_norm:} Uncertainty on `flux\_norm'
\item {\it flux\_tac:} Telluric absorption and instrument response-corrected flux density
\item {\it err\_tac:} Uncertainty on `flux\_tac'
\item {\it flux\_irc:} Instrument response-corrected flux density
\item {\it err\_irc:} Uncertainty on `flux\_icr'
\item {\it flux\_raw:} Raw flux density; output of the HERMES pipeline
\item {\it err\_raw:} Uncertainty on `flux\_raw'
\end{enumerate}
\end{samepage}

The wavelengths are expressed in angstroms, with an equidistant sampling in natural logarithm of the wavelength (of about $5.2 10^{-6}$), which is the baseline for the HERMES spectrograph, in order to facilitate velocity corrections without re-binning the data. The WCS keywords in the primary header describe that logarithmic wavelength scale.

Except in their normalised version, the spectra were scaled to an average flux of 1.0 over the wavelength range between 5400 and 5600\,\AA. The barycentric correction was applied to all spectra, allowing for direct measurement of the stellar radial velocity. The solar spectrum (obsid 884752) is the only one that has been corrected further, based on the radial velocity of Ganymede. 

All spectra obey the same format, including those which were not normalised. In the latter cases, the corresponding columns are filled with NaNs. Beyond 8950\,\AA, all normalised spectra are filled with NaNs. Two small wavelength ranges not covered by the HERMES spectrograph are also padded with NaNs in all spectra: $\sim$\,3\AA\ around 8578\,\AA\ and $\sim$\,8\AA\ around 8790\,\AA.

In the primary header, {\sc normlzd} is set to `True' if the spectrum was normalised, and the keywords {\sc `dscore', `dscmad', `degree',} and {\sc `it'} refer to the selected solution for the normalisation and its quality (see text). The following keywords describe the instrument response correction (linking to the corresponding meta-data file):
\begin{samepage}
\begin{enumerate}
\item {\sc unseq:} Unique sequence number = obsid
\item {\sc stdname:} Calibration target used for the instrument response correction
\item {\sc stdunseq:} Observation identifier of the calibration measurement
\item {\sc obsnight:} Observation date of {\sc unseq}
\item {\sc rspnight:} Observation date of {\sc stdunseq}
\item {\sc stdnight:} Boolean --- true if {\sc unseq} and {\sc stdunseq} were acquired the same night (i.e. if {\sc obsnight = rspnight})
\item {\sc stdstar:} Boolean --- true for the calibration spectra; false for the primary targets
\end{enumerate}
\end{samepage}
The format for the dates is yyyymmdd.0, and corresponds to the evening, applied to the entire night.

\subsection{Calibration spectra}

This dataset contains all the calibration spectra used to derive the instrument response on any given night of the programme. The spectra are presented in a similar format as presented in Section~\ref{sec:library} except that the binary table is limited to the two wavelength related columns plus the `flux\_raw' and `err\_raw' columns.

\subsection{Theoretical models}

This dataset gathers the theoretical models of the 15 calibration stars presented in Table~\ref{tab:P2} and used for the instrumental response corrections. The models are presented in ascii with two columns, namely, wavelength (in \AA) and  flux density (in erg~s$^{-1}$~cm$^{-2}$~\AA$^{-1}$).

\subsection{Meta-data}

The meta-data files are human-readable, pipe-separated ascii files with a heading and a trailing `|' character (written in `fixed\_width' format with astropy.io.ascii).

\subsubsection{Meta-data: spectral library}

The file melchiors\_meta\_spectra.txt provides the main parameters describing the spectra and the corresponding targets (the files have different names in CDS, but the correspondence is trivial). It contains the following fields:
\begin{enumerate}
\item {\it obsid:} Unique observation identifier.
\item {\it \sc target:} Target name. This corresponds to the HD number, if present,  and otherwise to the Hipparcos identifier or the Simbad name taken from the `starname' entry.  
\item {\it starname:} Simbad name of the target. 
\item {\it HD:} HD target identifier.  
\item {\it HIP:} HIP target identifier.  
\item {\it V:} Apparent V magnitude. 
\item {\it BV:} B-V colour index (unavailable values are set to 90).
\item {\it stype:} Spectral type. 
\item {\it comments:} Simbad code for the object type. 
\item {\it ra:} Right ascension (J2000). 
\item {\it dec:} Declination (J2000).
\item {\it pmra:} Proper motion in right ascension. 
\item {\it pmdec:} Proper motion in declination. 
\item {\it airmass:} Airmass. 
\item {\it exptime:} Exposure time. 
\item {\it usn, bsn, vsn, rsn, isn:} Approximate S/N in the U, B, V, R, and I bands.
\item {\it bvcor:} Barycentric velocity correction applied. 
\item {\it bjd:} Acquisition date (julian).
\item {\it date-avg:} Acquisition date.
\item {\it filename:} Name of the file containing the spectrum.
\item {\it stdnight:} {\sc stdnight} fits keyword. 
\item {\it stype\_ref:} Reference for the spectral type.
\item {\it dscore:} Normalisation quality index, in [0.00 - 0.24].
\end{enumerate}

Most of the fields are either filled from the fits headers or from a Simbad query. We refer to the text for more details. For the solar spectrum (obsid 884752), some of the meta-data are related to Ganymede and not the Sun itself (coordinates, barycentric correction).

\subsubsection{Meta-data: calibration spectra}

The file melchiors\_calibration\_match.txt provides the links between the spectral library and the calibration observations used for the instrument response correction. The information provided in this file was also reflected in the primary header of the spectra. This file contains the following five columns, obsid, cal\_obsid, cal\_name, night, and cal\_night, which respectively correspond to the following fits header entries, {\sc unseq, stdunseq, stdname, obsnight}, and {\sc rspnight} (see above). 

\subsubsection{Meta-data: Gaia-DR3}
\label{appendix:gaia}
The file melchiors\_gaiadr3.txt contains the result of the cross-matching of our dataset with the table gaia\_source of Gaia-DR3 using a cone search of 3~arcsec for 1592 MELCHIORS stars. It reports all fields from the Gaia-DR3, and we therefore direct the reader to \href{https://gea.esac.esa.int/archive/documentation/GDR3/Gaia\_archive/chap\_datamodel/sec\_dm\_main\_source\_catalogue/ssec\_dm\_gaia\_source.html}{the corresponding Gaia documentation}. The  field {\sc target} is common with melchiors\_meta\_spectra.txt.
Concerning binary stars, the field non\_single\_star identifies them with values of 1, 2, 3, and 4 for astrometric, spectroscopic, astro+spectroscopic, and eclipsing binaries, respectively.

\subsection{Quick-look}
A quick-look image is available for every spectrum in the library in .png format. Each image shows the entire spectrum before and after telluric absorption correction as well as a zoom in $H\alpha$ before and after normalisation. These images are accessible individually on the website and are included in the bulk download of the auxiliary data.

\end{appendix}

\end{document}